\documentclass[prd,twocolumn,floatfix,amsmath,nofootinbib,amssymb,floatfix]{revtex4}
\usepackage{graphicx,color,dcolumn,booktabs,bm}
\usepackage{longtable,lscape}
\usepackage{pdfpages}
\usepackage{txfonts}
\usepackage{overpic}
\usepackage{amssymb}
\usepackage{makecell}
\usepackage{indentfirst}
\usepackage{feynmf}   
\usepackage{slashed}  
\usepackage{cases}
\usepackage{color}
\usepackage{multirow}
\usepackage{threeparttable}
\usepackage{enumerate}
\usepackage{subfigure}
\usepackage{diagbox}
\usepackage{siunitx}
\usepackage{mathrsfs}
\usepackage{cancel}
\usepackage{float}
\usepackage[colorlinks,
citecolor=blue,
anchorcolor=red,
menucolor=red,
linkcolor=red,
filecolor=red,
runcolor=red,
urlcolor=blue,
frenchlinks=red]{hyperref}

\usepackage{epstopdf}
\usepackage{amsmath}
\usepackage{graphicx}
\usepackage{adjustbox}
\usepackage{dutchcal}
\usepackage{array}
\usepackage{booktabs}
\usepackage{makecell}
\usepackage{placeins}
\usepackage{textcomp}
\usepackage{amsmath}
\DeclareRobustCommand{\perthousand}{%
	\ifmmode
	\text{\textperthousand}%
	\else
	\textperthousand
	\fi}
\begin{document}
	\title{Analysis of the strong vertices of hadronic molecules $DK$, $D^*K$, $DK^*$ and their bottom analogs}
	\author{Ze Zhou$^{1}$}
	\author{Guo-Liang Yu$^{1,2}$}
	\email{yuguoliang2011@163.com}
	\author{Zhi-Gang Wang$^{1,2}$}
	\email{zgwang@aliyun.com}
	\author{Jie Lu$^{3}$}
	\affiliation{$^1$ Department of Mathematics and Physics, North China
		Electric Power University, Baoding 071003, People's Republic of
		China\\$^2$ Hebei Key Laboratory of Physics and Energy Technology, North China Electric Power University, Baoding 071000, China\\$^3$  School of Physics, Southeast University, Nanjing 210094, People’s Republic of China}
	\date{\today}
\begin{abstract}
In this work, we analyze the strong vertices of hadronic molecules $DK$, $D^*K$, $DK^*$ and their bottom analogs within the framework of three-point QCD sum rules. The coupling between interpolating currents and low spin particles is considered in the phenomenological side, and the vacuum condensates $\left\langle \bar qq \right\rangle ,\left\langle g_s^2GG \right\rangle ,\left\langle \bar qg_s\sigma Gq \right\rangle ,\left\langle g_s^3GGG \right\rangle ,{\left\langle \bar qq \right\rangle ^2}$ are included in the QCD side. As an application of strong coupling constants, we also obtain the partial decay widths of these states, where $\Gamma_{T_{DK}\rightarrow D_s\pi}=10.3_{-2.9}^{+3.0}$ KeV,
 $\Gamma_{T_{D^*K}\rightarrow D_s^*\pi}=24.8_{-5.4}^{+5.5}$ KeV, $\Gamma_{T_{BK}\rightarrow BK}=101_{-21}^{+36}$ MeV and
$\Gamma_{T_{B^*K}\rightarrow B^*K}=142_{-25}^{+52}$ MeV. It is shown that the results of $\Gamma_{T_{DK}\rightarrow D_s\pi}$ and $\Gamma_{T_{D^*K}\rightarrow D_s^*\pi}$ are compatible with the experimental data of $D_{s0}^*(2317)$ and $D_{s1}(2460)$.
\end{abstract}
\maketitle

\section{Introduction}\label{sec1}
Over the past two decades, numerous new hadronic states have been discovered like bamboo shoots after a spring rain, which have aroused the great interest of theoretical physicists. A representative of these hadronic states is $D_{s1}(2536)$ which was observed in 1987 by analyzing the $D_s^*\gamma$ invariant mass spectrum in the $\bar \nu N$ scattering process \cite{Asratian:1987rb}. Later in 1989, ARGUS collaboration also observed this state in the $e^+e^-$ collision at the DESY storage ring \cite{ARGUS:1989zue}. In 2003, another charm-strange state $D_{s0}^*(2317)$ was also observed in the $D_s^+\pi^0$ mass distribution in $e^+e^-$ annihilation process by BABAR collaboration \cite{BaBar:2003oey}. The experimental data indicates that its mass is located near $2.32$ GeV and its decay width is very narrow. In the subsequent experimental research to confirm this state, another narrow state $D_{s1}(2460)$ was discovered in the $D_s^*\pi^0$ channel by CLEO collaboration \cite{CLEO:2003ggt}. Since then, the existence of these charm-strange states was confirmed by many experiments \cite{Belle:2003guh, BaBar:2004yux, BaBar:2006eep, CLEO:1993nxj, Belle:2019qoi, LHCb:2023eig}, and their masses and decay widths were determined as,	
\begin{eqnarray}
\notag
&&m_{D_{s0}^*}(2317)=2317.7\pm0.5 \ \mathrm{MeV}\\
\notag
&&\Gamma_{D_{s0}^*(2317)}<3.8\ \mathrm{MeV}\\
\notag
&&m_{D_{s1}}(2460)=2459.5\pm 0.6\ \mathrm{MeV}\\
\notag
&&\Gamma_{D_{s1}(2460)}<3.5\ \mathrm{MeV}\\
\notag
&&m_{D_{s1}}(2536)=2535.11\pm 0.06\ \mathrm{MeV}\\
\notag
&&\Gamma_{D_{s1}(2536)}=0.9\pm 0.05\ \mathrm{MeV}
\end{eqnarray}

The internal quark composition of these states was initially explained as $c\bar{s}$ component. However, the experimental masses of these states are clearly lower than the predictions of Goldfrey-Isgur (GI) model \cite{Godfrey:1985xj}. This phenomenon inspired widespread attention from theoretical physicists about the nature of these charm-strange states. Theorists proposed various interpretations about their structures, such as charm-strange meson \cite{Bardeen:2003kt, Deandrea:2003gb, Dai:2003yg, Sadzikowski:2003jy, Cahn:2003cw, Hwang:2004cd, Simonov:2004ar, Cheng:2014bca, Song:2015nia, Cheng:2017oqh, Luo:2021dvj, Zhou:2020moj, Alhakami:2016zqx} and tetraquark states \cite{Nielsen:2005ia, Wang:2006uba, Dmitrasinovic:2005gc, Cheng:2003kg, Terasaki:2003qa, Dmitrasinovic:2004cu, Kim:2005gt, Zhang:2018mnm}. Besides, because these states locate near the meson-meson thresholds, they can also be explained as the hadronic molecules \cite{vanBeveren:2003kd, Chen:2004dy, Liu:2022dmm, Kolomeitsev:2003ac, Gamermann:2007fi, Guo:2006fu, Kong:2021ohg, Barnes:2003dj, Gamermann:2006nm, Guo:2006rp, Guo:2009ct, Xie:2010zza, Wu:2011yb, Guo:2015dha, Du:2017ttu, Guo:2018tjx, Albaladejo:2018mhb, Wu:2019vsy, Wang:2012bu, Huang:2021fdt}. In our previous work \cite{Zhou:2025yjb}, the masses and pole residues of $DK$, $D^*K$ and $DK^*$ molecular states and their flavor partners $BK$, $B^*K$ and $BK^*$ were analyzed, and the results support the interpretation of treating $D_{s0}^*(2317)$, $D_{s1}(2460)$ and $D_{s1}(2536)$ as $DK$, $D^*K$ and $DK^*$ molecules. The predicted mass of $BK^*$ is consistent well with $B_{sJ}(6158)$ measured by LHCb Collaboration \cite{LHCb:2020pet}. As a continuation of this work, we will analyze the strong coupling constants of these charm- and bottom-strange states in the present work by supporting them as hadronic molecules, which are related to their decay and production processes.
	
Since the masses of $D_{s0}^*(2317)$ and $D_{s1}(2460)$ are lower than the thresholds of $DK$ and $D^*K$, respectively, the decay processes $D_{s0}^*(2317) \to DK$ and ${D_{s1}}(2460) \to D^*K$ are kinematically forbidden. Their decay processes can be explained through the isospin violation process, which includes two steps, $D_{s0}^*(2317) \to {D_s}\eta  \to {D_s}\pi^0$ and ${D_{s1}}(2460) \to D_s^*\eta  \to D_s^*\pi^0$. The second step is driven by the $\eta - \pi^0$ mixing mechanism triggered by the masses difference between $u$ and $d$ quark. The value of $\eta - \pi^0$ transition matrix is quite small according to Dashen's theorem \cite{Dashen:1969eg}, $t_{\eta \pi}=\left\langle {\pi^0} \right|H\left| \eta \right\rangle $ = $-$0.003 GeV$^2$. This mechanism provides a good explanation for their narrow widths in experiments. Based on this mechanism, there already have been many studies which analyzed the $\pi$ decay process \cite{Faessler:2007gv, Han:2023wqq, Fu:2021wde, Cleven:2014oka}. Supporting $D_{s1}(2536)$ as a $DK^{*}$ molecular state, we mainly focus on its $D^*K$ decay channel because it occupies a large branching ratio in the total width.
	
\begin{figure}
		\centering
		\includegraphics[width=8cm, trim=80 480 60 50, clip]{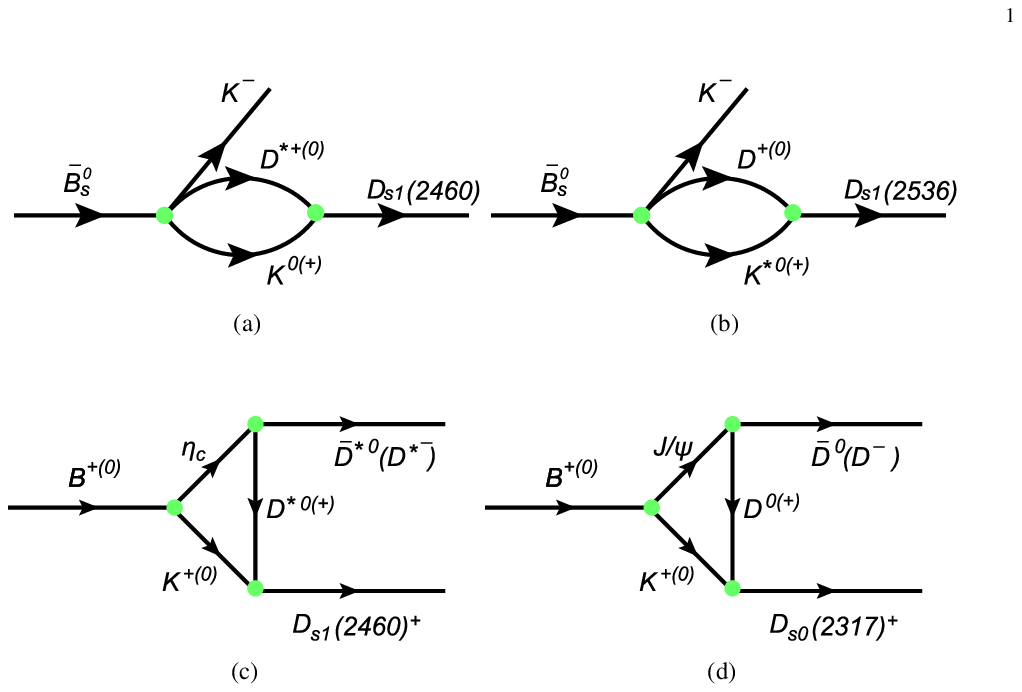}
		\caption{The production processes of $D_{s0}^*(2317)$, $D_{s1}(2460)$ and $D_{s1}(2536)$}
		\label{generation process 2317 2460 2536}
\end{figure}

In addition to the decay processes of these charm-strange states, their production processes are also important to understand their inner structures. For example, the decay processes of $B \to D_{s0}^*(2317)/D_{s1}(2460)\bar{D}^{(*)}$ and $\bar{B}_s^0 \to D_{s1}(2460)^+/D_{s1}(2536)^+K^-$ were investigated in the molecular picture in Ref \cite{Faessler:2007cu, Liu:2022dmm, Lin:2024hys}. These decay processes can be explained by the feynman diagrams shown in Fig. \ref{generation process 2317 2460 2536}. In our previous work, strong vertices $J/\psi DD$ and $\eta_c D^*D^*$ which are related to Fig. \ref{generation process 2317 2460 2536} (c) and (d) were studied with the three-point QCD sum rules (QCDSR) \cite{Lu:2023gmd}. In the present work, we will analyze the strong vertices $D_{s0}^*(2317)DK$, $D_{s1}(2460)D^*K$ and $D_{s1}(2536)DK^*$ by treating these states as molecular states.

Motivated by these above analyses, we will study the strong coupling constants of $T_{DK}DK$, $T_{D^{*}K}{D^*}K$, $T_{DK^{*}}D{K^*}$, $T_{DK^{*}}D^*K$, $T_{DK}D_s\eta$ and $T_{D^{*}K}D_s^*\eta$. As the partners of open charm molecular states, the strong vertices of hadronic molecules $BK$, $B^*K$ and $BK^*$ will also be analyzed in the present work, which is helpful for searching for these states in experiments in the future. This article is arranged as follows: After the introduction in Sec. \ref{sec1}, we analyze the strong vertices in the framework of three-point QCDSR in Sec. \ref{sec2}, and the vacuum condensates up to dimension 6 at the QCD side are considered. The numerical results and discussions are given in Sec. \ref{sec3}. In Sec. \ref{sec4}, the partial decays of these molecular states are calculated by using obtained strong coupling constants. Sec. \ref{sec5} is reserved for a short conclusion.
\section{Three-point QCD sum rules for coupling constants}\label{sec2}
In order to obtain strong coupling constants which we are interested in, we firstly write down the following three-point correlation functions,
\begin{eqnarray}
\notag
\Pi^{\texttt{phy}1}(p,q)=&&i^2\int d^4xd^4ye^{ip\cdot x}e^{iq\cdot y}\\
\notag
&&\times\left\langle0\right|T[J^{D[B]}(x)J^K(y)J^{T_{D[B]K}\dag}(0)]\left|0\right\rangle\\
\notag
\Pi_{\mu\nu}^{\texttt{phy}2}(p,q)=&&i^2\int d^4xd^4ye^{ip\cdot x}e^{iq\cdot y}\\
\notag
&&\times\left\langle0\right|T[J_\mu^{D^*[B^*]}(x)J^K(y)J_\nu^{T_{D^*[B^*]K}\dag}(0)]\left|0\right\rangle\\
\notag
\Pi_{\mu\nu}^{\texttt{phy}3}(p,q)=&&i^2\int d^4xd^4ye^{ip\cdot x}e^{iq\cdot y}\\
\notag
&&\times\left\langle0\right|T[J^{D[B]}(x)J_\mu^{K^*}(y)J_\nu^{T_{D[B]K^*}\dag}(0)]\left|0\right\rangle
\end{eqnarray}
\begin{eqnarray}
\notag
\Pi_{\mu\nu}^{\texttt{phy}4}(p,q)=&&i^2\int d^4xd^4ye^{ip\cdot x}e^{iq\cdot y}\\
\notag
&&\times\left\langle0\right|T[J_\mu^{D^*[B^*]}(x)J^K(y)J_\nu^{T_{D[B]K^*}\dag}(0)]\left|0\right\rangle\\
\notag
\Pi_\rho^{\texttt{phy}5}(p,q)=&&i^2\int d^4xd^4ye^{ip\cdot x}e^{iq\cdot y}\\
\notag
&&\times\left\langle0\right|T[J^{D_s}(x)J_{\rho}^{\tilde\eta}(y)J^{T_{DK}\dag}(0)]\left|0\right\rangle\\
\notag
\Pi_{\rho\mu\nu}^{\texttt{phy}6}(p,q)=&&i^2\int d^4xd^4ye^{ip\cdot x}e^{iq\cdot y}\\
&&\times\left\langle0\right|T[J_\mu^{D_s^*}(x)J_{\rho}^{\tilde\eta}(y)J_\nu^{T_{D^*K}\dag}(0)]\left|0\right\rangle
\end{eqnarray}
where $T$ is the time ordered product, and $T_{AB}$ represents the tetraquark molecular states composed of $A$ and $B$. $J$ is the interpolating current that can couple to studied state. The currents of these states can be written as,
\begin{eqnarray}\label{current of four quark states1}
\notag
J^{D[B]}=&&\bar u^{n}i\gamma_5Q^{n} \\
\notag  J_\mu^{D^*[B^*]}=&&\bar u^{n}\gamma_\mu Q^{n}\\
\notag
J^{D_s}=&&\bar s^{n}i\gamma_5c^{n} \\
\notag  J_\mu^{D_s^*}=&&\bar s^{n} \gamma_\mu c^ {n}\\
\notag
J^K=&&\bar s^{n}i\gamma_5u^{n} \\
 J_\mu^{K^*}=&&\bar s^{n}\gamma_ \mu u^{n}
\end{eqnarray}
and
\begin{eqnarray}\label{current of four quark states2}
\notag&&J_\rho^{\bar\eta}=\frac{1}{\sqrt6}[\bar u^{n}\gamma_\rho\gamma_5u^{n}+\bar d^{n} \gamma_\rho\gamma_5d^{n}-2\bar s^{n}\gamma_\rho\gamma_5s^{n}]\\
\notag
&&J_\nu^{T_{D^*[B^*]K}}=\frac{1}{\sqrt2}[\bar s^mi\gamma_5u^m\bar u^n\gamma_\nu Q^n +\bar s^mi\gamma_5d^m\bar d^n\gamma_\nu Q^n]\\
\notag
&&J_\nu^{T_{D[B]K^*}}=\frac{1}{\sqrt2}[\bar s^m\gamma_\nu u^m\bar u^ni\gamma_5Q^n +\bar s^m\gamma_\nu d^m\bar d^ni\gamma_5Q^n]\\
&&J^{T_{D[B]K}}=\frac{1}{\sqrt2}[\bar s^mi\gamma_5u^m\bar u^ni\gamma_5Q^n+\bar s^mi \gamma_5d^m\bar d^ni\gamma_5Q^n]
\end{eqnarray}
where $m$ and $n$ are color indices, and $Q$ stands for $c$ and $b$ quark fields.

In the framework of QCD sum rules \cite{Shifman:1978by, Shifman:1978bx}, the correlation function will be treated both in the hadron and quark levels which are called the phenomenological and QCD sides. In the hadron level, these currents can couple to hadronic states that we studied. In the quark level, it will be dealt by operator product expansion (OPE), where the long distance quark-gluon interactions are absorbed in vacuum condensates, and the short distance interactions are absorbed in Wilson coefficients. Finally, according to the quark-hadron duality, the strong coupling constants can be extracted and the sum rules for coupling constants are obtained.
	
\subsection{The phenomenological side}\label{The phenomenological side}
	
In the phenomenological side, the complete sets of intermediate hadronic states that can couple to the current operators are inserted into correlation functions. After performing integration in coordinate space, we isolate the contribution of ground state. Then, the three-point correlation functions can be represented as following forms,
\begin{widetext}
	
\begin{eqnarray}
\notag
\Pi^{\texttt{phy}1}(p,q)=&&\frac{\left\langle0\right|J^{D[B]}(0)\left|D[B](p)\right\rangle\left\langle0\right|J^K(0)\left|K(q)\right\rangle\left\langle D[B](p)K(q)
\mathrel{\left|\vphantom{D[B](p)K(q)T_{D[B]K}(p')}\right.\kern-\nulldelimiterspace}
T_{D[B]K}(p')\right\rangle\left\langle T_{D[B]K}(p')\right|J^{T_{D[B]K}\dag}(0) \left|0\right\rangle}{(p^2-m_{D[B]}^2)(q^2-m_K^2)(p{'^2}-m_{T_{D[B]K}}^2)}+\cdot\cdot\cdot\\
\notag
&&=\Pi_1^{\texttt{phy}1}\\
\notag
\Pi_{\mu\nu}^{\texttt{phy}2}(p,q)=&&\frac{\left\langle0\right|J_\mu^{D^*[B^*]}(0)\left|D^*[B^*](p)\right\rangle\left\langle0\right|J^K(0)\left|K(q)\right\rangle\left\langle D^*[B^*](p)K(q)\mathrel{\left|\vphantom {D^*[B^*](p)K(q)T_{D^*[B^*]K}(p')}
\right.\kern-\nulldelimiterspace}T_{D^*[B^*]K}(p')\right\rangle\left\langle T_{D^*[B^*]K}(p')\right|J_\nu^{T_{D^*[B^*]K}\dag}(0)\left|0\right\rangle}{(p^2-m_{D^*[B^*]}^2)(q^2-m_K^2)(p{'^2}-m_{T_{D^*[B^*]K}}^2)}\\
\notag
&&+\cdot\cdot\cdot\\
\notag
=&&\Pi_1^{\texttt{phy}2}g_{\mu\nu}+\Pi_2^{\texttt{phy}2}p_\mu p_\nu+\Pi_3^{\texttt {phy}2}p^\prime_\mu p^\prime_\nu+\Pi_4^{\texttt{phy}2}p_\mu p^\prime_\nu\\
\notag
\Pi_{\mu\nu}^{\texttt{phy}3}(p,q)=&&\frac{\left\langle0\right|J^{D[B]}(0)\left|D[B](p)\right\rangle\left\langle0\right|J_\mu^{K^*}(0)\left|K^*(q)\right\rangle\left\langle D[B](p)K^*(q)\mathrel{\left|\vphantom{D[B](p)K^*(q)T_{D[B]K^*}(p')}\right. \kern-\nulldelimiterspace}T_{D[B]K^*}(p')\right\rangle\left\langle T_{D[B]K^*} (p')\right|J_\nu^{T_{D[B]K^*}\dag}(0)\left|0\right\rangle}{(p^2-m_{D[B]}^2)(q^2-m_{K^*}^2)(p{'^2}-m_{T_{D[B]K^*}}^2)}\\
\notag
&&+\cdot\cdot\cdot\\
\notag
=&&\Pi_1^{\texttt{phy}3}g_{\mu\nu}+\Pi_2^{\texttt{phy}3}q_\mu q_\nu+\Pi_3^{\texttt {phy}3}p^\prime_\mu p^\prime_\nu+\Pi_4^{\texttt{phy}3}q_\mu p^\prime_\nu\\
\notag
\Pi_{\mu\nu}^{\texttt{phy}4}(p,q)=&&\frac{\left\langle0\right|J^{D^*[B^*]}(0)\left|D^*[B^*](p)\right\rangle\left\langle0\right|J^K(0)\left|K(q)\right\rangle\left\langle D^*[B^*](p)K(q)\mathrel{\left|\vphantom{D^*[B^*](p)K(q)T_{D[B]K^*}(p')}\right. \kern-\nulldelimiterspace}T_{D[B]K^*}(p') \right\rangle\left\langle T_{D[B]K^*}(p') \right|J^{T_{D[B]K^*}\dag}(0)\left|0\right\rangle}{(p^2-m_{D^*[B^*]}^2)(q^2-m_K^2)(p{'^2}-m_{T_{D[B]K^*}}^2)}\\
\notag
&&+\cdot\cdot\cdot\\
\notag
=&&\Pi_1^{\texttt{phy}4}g_{\mu\nu}+\Pi_2^{\texttt{phy}4}p_\mu p_\nu+\Pi_3^{\texttt {phy}4}p^\prime_\mu p^\prime_\nu+\Pi_4^{\texttt{phy}4}p_\mu p^\prime_\nu\\
\notag
\Pi^{\texttt{phy}5}(p,q)=&&q^\rho\frac{\left\langle0\right|J^{D_s}(0)\left| D_s (p)\right\rangle\left\langle0\right|J_\rho^{\tilde\eta}(0)\left|\eta(q)\right\rangle\left\langle D_s(p)\eta (q)\mathrel{\left|{\vphantom {D_s(p)\eta(q){T_{DK}(p')}}}
\right.\kern-\nulldelimiterspace}T_{DK}(p')\right\rangle\left\langle T_{DK}(p') \right|J^{T_{DK}\dag}(0)\left|0\right\rangle}{(p^2-m_{D_s}^2)(q^2-m_\eta^2)(p{'^2}-m_{T_{DK}}^2)}+\cdot\cdot\cdot\\
\notag
=&&\Pi_1^{\texttt{phy}5}\\
\notag
\Pi_{\lambda\sigma}^{\texttt{phy}6}(p,q)=&&\frac{\left\langle0\right|J_\mu^{D_s^*}(0)\left|D_s^*(p)\right\rangle\left\langle0\right|J_\rho^{\tilde\eta}(0)\left|\eta(q)\right\rangle\left\langle D_s^*(p)\eta (q)\mathrel{\left|\vphantom{D_s^*(p)\eta(q) T_{D^*K}(p')}\right.\kern-\nulldelimiterspace}T_{D^*K}(p')\right\rangle\left\langle T_{D^*K}(p')\right|J_\nu^{T_{D^*K}\dag}(0)\left|0\right\rangle}{(p^2-m_{D_s^*}^2)(q^2-m_\eta^2)(p{'^2}-m_{T_{D^*K}}^2)}\\
\notag
&&\times q^\rho(g^{\lambda\mu}-\frac{p^\lambda p^\mu}{p^2})(g^{\sigma\nu} -\frac{p {'^\sigma }p{'^\nu }}{p{'^2}})+\cdot\cdot\cdot\\
=&&\Pi_1^{\texttt{phy}6}g_{\lambda\sigma}+\Pi_2^{\texttt{phy}6}p_\lambda p_\sigma+ \Pi_3^{\texttt{phy}6}p_\lambda p^\prime_\sigma+\Pi_4^{\texttt{phy}6}p^\prime_ \lambda p^\prime_ \sigma
\end{eqnarray}
\end{widetext}
where projection operator $q_{\rho}$ eliminates the coupling of axial vector current $J_\rho^{\bar \eta}$ with axial vector particle, and $(g^{\lambda \mu } - \frac{p^\lambda p^\mu}{p^2})(g^{\sigma \nu} - \frac{{p'}^\sigma {p'}^\nu }{{p'}^2})$ eliminates the coupling of currents $J_\mu^{D_s^*}(x)$ and $J_\nu ^{T_{D^*K}}$ with low spin states. $\Pi^{\texttt{phy}j}_{i}$ is scalar invariant amplitude. The ellipsis denotes the contributions of excited and continuum states. The matrix elements in these equations can be defined as,
\begin{eqnarray}\label{matrix element}
\notag
&&\left\langle D_s(p)\eta (q)\mathrel{\left|\vphantom {D_s(p)\eta (q)T_{DK}(p')} \right.\kern-\nulldelimiterspace}T_{DK}(p')\right\rangle=-G_{T_{DK}D_s\eta}\\
\notag
&&\left\langle D_s^*(p)\eta(q)\mathrel{\left|\vphantom{D_s^*(p)\eta(q)T_{D^*K}(p')} \right.\kern-\nulldelimiterspace}T_{D^*K}(p')\right\rangle=G_{T_{D^*K}D_s^*\eta}\xi\cdot\varepsilon\\
\notag
&&\left\langle D[B](p)K(q)\mathrel{\left|\vphantom{D[B](p)K(q)T_{D[B]K}(p')}\right. \kern-\nulldelimiterspace}T_{D[B]K}(p')\right\rangle=-G_{T_{D[B]K}D[B]K}\\
\notag
&&\left \langle D^*[B^*](p)K(q)\mathrel{\left|\vphantom{D^*[B^*](p)K(q)T_{D[B]K^*} (p')}\right.\kern-\nulldelimiterspace}T_{D[B]K^*}(p')\right\rangle=G_{T_{D^*[B^*]K}D[B]K^*}\zeta\cdot\varepsilon\\
\notag
&&\left \langle D^*[B^*](p)K(q)\mathrel{\left|\vphantom{D^*[B^*](p)K(q)T_{D^*[B^*] K}(p')} \right.\kern-\nulldelimiterspace}T_{D^*[B^*]K}(p')\right\rangle=G_{T_{D^*[B ^*]K}D^*[B^*]K}\xi\cdot\varepsilon\\
&&\left\langle D[B](p)K^*(q)\mathrel{\left|\vphantom {D[B](p)K^*(q)T_{D[B]K^*}(p')}
\right.\kern-\nulldelimiterspace}T_{D[B]K^*}(p')\right\rangle=G_{T_{D[B]K^*}D[B]K^*}\zeta\cdot\varepsilon
\end{eqnarray}
and
\begin{eqnarray}
\notag	&&\left\langle0\right|J_\rho^{\tilde\eta}(0)\left|{\eta(q)}\right\rangle=if_\eta q_\rho\\
\notag
&&\left\langle0\right|J^K(0)\left|K(q)\right\rangle=\frac{f_Km_K^2}{m_u+m_s}\\
\notag
&&\left\langle0\right|J^{D_s}(0)\left|D_s(p)\right\rangle=\frac{f_{D_s}m_{D_s}^2}{m_c+m_s}\\
\notag
&&\left\langle0\right|J^{D[B]}(0)\left|D[B](p)\right\rangle=\frac{f_{D[B]}m_{D[B]}^2}{m_u+m_{c[b]}}\\
\notag
&&\left\langle0\right|J_\mu^{D_s^*}(0)\left|D_s^*(p)\right\rangle=f_{D_s^*}m_{D_s^*}{\xi_\mu}\\
\notag
&&\left\langle0\right|J_\mu^{K^*}(y)\left|K^*(q)\right\rangle=f_{K^*}m_{K^*}{\zeta _\mu}\\
\notag
&&\left\langle0\right|J_\mu^{D^*[B^*]}(0)\left|D^*[B^*](p)\right\rangle=f_{D^*[B^*]}m_{D^*[B^*]}{\xi_\mu}\\
\notag
&&\left\langle T_{D^*[B^*]K}(p')\right|J_\nu^{T_{D^*[B^*]K}\dag}(0)\left|0\right \rangle=f_{T_{D^*[B^*]K}}{\varepsilon_\nu}\\
\notag
&&\left\langle T_{D[B]K}(p')\right|J^{T_{D[B]K}\dag}(0)\left|0\right\rangle=f_{T_{D [B]K}}\\
&&\left\langle T_{D[B]K^*}(p')\right|J_\nu^{T_{D[B]K^*}\dag}(0)\left|0\right\rangle= f_{T_{D[B]K^*}}{\varepsilon_\nu}
\end{eqnarray}
where $\varepsilon$, $\zeta$ and $\xi$ are polarization vector of vector or axial-vector particles, and has following property,
\begin{eqnarray}\label{polerazation sum}
\mathop\sum\limits_\lambda\eta_\mu^*(\lambda,p)\eta_\nu(\lambda,p)=-g_{\mu\nu}+\frac{p_\mu p_\nu}{p^2},(\eta=\varepsilon,\zeta,\xi)
\end{eqnarray}
In this work, we choose scalar invariant amplitude $\Pi_1^{\texttt{phy}j}$$(j=1\sim6)$ to analyze the strong coupling constants. In the phenomenological side, the correlation function can be expressed with the hadronic spectral densities $\rho^\texttt{phy}$ through the triple dispersion relation,
\begin{eqnarray}
\notag
&&\Pi^\texttt{phy}(p{'^2},p^2,q^2)\\
&&=\int\limits_{s{'_t}}^\infty ds'\int\limits_ {s_t}^\infty ds \int\limits_{u_t}^\infty du\frac{\rho^\texttt{phy}(s',s,u)}{(s'-p'^2)(s-p^2) (u-q^2)}
\end{eqnarray}
with
\begin{eqnarray}
\notag
&&\rho^\texttt{phy}(s',s,u)\\
&&=\lim_{\varepsilon_{1/2/3}\to0}\frac{\mathop{\rm Im} \nolimits_{s'}\mathop{\rm Im}\nolimits_s\mathop{\rm Im}\nolimits_u\Pi(s'+i \varepsilon_1,s +i\varepsilon_2,u+i\varepsilon_3)}{\pi^3}
\end{eqnarray}
where $s'_t$, $s_t$ and $u_t$ are the kinematic limit. However, we can only conduct dual dispersion relation in the QCD side,
\begin{eqnarray}
\Pi^\texttt{QCD}(p{'^2},p^2,q^2)=\int\limits_{s_t}^\infty ds\int\limits_{u_t}^ \infty du\frac{\rho^\texttt{QCD}(p{'^2},s,u)}{(s-p^2)(u-q^2)}
\end{eqnarray}
because \cite{Wang:2024qqa},
\begin{eqnarray}
\mathop{\rm lim}_{\varepsilon_1\to0}\mathop{\mathop{\rm Im}\nolimits_{s'}}\Pi^{\texttt{QCD}}(s'+i\varepsilon_1,p^{2},q^{2})=0
\end{eqnarray}
It is clear that the phenomenological side and QCD side of the correlation function are not equivalent with each other. There are two methods to deal with this issue. The first one which we call method I is conducting the integral over $ds'$ in the phenomenological side, and introducing a parameter $C_i$ to parameterize the contributions of excited and continuum states in the $s'$ channel \cite{Wang:2017lot, Wang:2019iaa, Wang:2025sic}. In this case, $\Pi_1^{\texttt{phy}j}(j=1\sim6)$ can be represented as the following forms,
\begin{widetext}
\begin{eqnarray}
\notag&&\Pi_1^{\texttt{phy}1}(p{'^2},p^2,q^2)=\frac{f_{D[B]}m_{D[B]}^2f_Km_K^2G_{T_{D[B]K}D[B]K}f_{T_{D[B]K}}}{(m_u+m_{c[b]})(m_u+m_s)(p^2-m_{D[B]}^2)(q^2-m_K^2)(p{'^2}-m_{T_{D[B]K}}^2)}+\frac{C_{1[5]}}{(p^2 - m_{D[B]}^2)(q^2-m_K^2)}+\cdot\cdot\cdot\\
\notag&&\Pi_1^{\texttt{phy}2}(p{'^2},p^2,q^2)=\frac{f_{D^*[B^*]}m_{D^*[B^*]}f_Km_K^2G_{T_{D^*[B^*]K}D^*[B^*]K}f_{T_{D^*[B^*]K}}}{(m_u+m_s)(p^2-m_{D^*[B^*]}^2)(q^2-m_K^2)(p{'^2}-m_{T_{D^*[B^*]K}^2})}+\frac{C_{2[6]}}{(p^2-m_{D^*[B^*]}^2)(q^2-m_K^2)}+\cdot\cdot\cdot\\
\notag&&\Pi_1^{\texttt{phy}3}(p{'^2},p^2,q^2)=\frac{f_{D[B]}m_{D[B]}^2f_{K^*}m_{K^*}G_{T_{D[B]K^*}D[B]K^*}f_{T_{D[B]K^*}}}{(m_u+m_{c[b]})(p^2-m_{D[B]}^2)(q^2-m_{K^*}^2)(p{'^2}-m_{T_{D[B]K^*}}^2)}+\frac{C_{3[7]}}{(p^2-m_{D[B]}^2)(q^2-m_{K^*}^2)}+\cdot\cdot\cdot\\
\notag&&\Pi_1^{\texttt{phy}4}(p{'^2},p^2,q^2)=\frac{f_{D^*[B^*]}m_{D^*[B^*]}f_Km_K^2G_{T_{D[B]K^*}D^*[B^*]K}f_{T_{D[B]K^*}}}{(m_u+m_s)(p^2-m_{D^*[B^*]}^2)(q^2-m_K^2)(p{'^2}-m_{T_{D[B]K^*}}^2)}+\frac{C_{4[8]}}{(p^2-m_{D^*[B^*]}^2)(q^2-m_K^2)}+\cdot\cdot\cdot\\
\notag&&\Pi_1^{\texttt{phy}5}(p{'^2},p^2,q^2)=\frac{f_{D_s}m_{D_s}^2if_\eta m_\eta^2 G_{T_{DK}D_s\eta}f_{T_{DK}}}{(m_c+m_s)(p^2-m_{D_s}^2)(q^2-m_\eta^2)(p{'^2}-m_{T_{DK}}^2)}+\frac{iC_9}{{(p^2-m_{D_s}^2)(q^2-m_\eta^2)}}+\cdot\cdot\cdot\\
&&\Pi_1^{\texttt{phy}6}(p{'^2},p^2,q^2) = \frac{f_{D_s^*}m_{D_s^*}if_\eta m_\eta ^2G_{T_{D^*K}D_s^*\eta}f_{T_{D^*K}}}{(p^2-m_{D_s^*}^2)(q^2-m_\eta^2)(p{'^2}-m_{T_{D^*K}}^2)}+\frac{iC_{10}}{(p^2-m_{D_s^*}^2)(q^2-m_\eta ^2)}+\cdot\cdot\cdot
\end{eqnarray}
\end{widetext}
where $C_i$ denote the contributions of excited and continuum states of initial particles, and they can be expressed as,
\begin{eqnarray}
C=\int\limits_{s'_0}^\infty ds'\frac{\rho^\texttt{phy}(s',m_2^2,m_3^2)}{(s'-{p'}^2)}
\end{eqnarray}
where $s'_0$ is the threshold parameter for isolating the ground state from the first excited state in the $p'^2$ channel. $m_2$ and $m_3$ are the masses of ground states in the channels $p^2$ and $q^2$, respectively. As the specific form of spectral density ${\rho (s',m_2^2,m_3^2)}$ is unknown, $C_i$ is an adjustable parameter to acquire reliable QCD sum rules.

Another approach which we call method II is to setting initial particle off-shell in deep space-like region by taking $P'^2=-p'^2$. Under the circumstances, we obtain the scalar invariant amplitude with variation of momentum $P'^2$ to avoid triple dispersion relation in the hadron side. It is noted that this method is quasi quark-hadron duality, as it ignores the contributions of excited and continuum states in $p'^2$ channel. However, as a good approximation method, it is still widely used to analyze the strong coupling constants. In the present work, we will use both method I and II to carry out the calculations, and will compare the results obtained with these two methods to verify the reliability of our results.

\subsection{The QCD side}
After contracting all the quark fields with wick's theorem, the correlation functions can be represented as the following forms in the QCD side,
\begin{widetext}
\begin{eqnarray}\label{contraction of TDK}
\notag
\Pi^{\texttt{QCD}1}(p,q)=&&-\frac{1}{\sqrt2}\int d^4xd^4ye^{ip\cdot x}e^{iq\cdot y} Tr[U^{nn'}(-x)\gamma_5Q^{n'n}(x)\gamma_5]Tr[S^{mm'}(-y)\gamma_5U^{m'm}(y)\gamma_5]\\
\notag
\Pi_{\mu\nu}^{\texttt{QCD}2}(p,q)=&&\frac{1}{\sqrt2}\int d^4xd^4ye^{ip\cdot x}e^{iq \cdot y}Tr[U^{nn'}(-x)\gamma_5Q^{n'n}(x)\gamma_5]Tr[S^{mm'}(-y)\gamma_\nu U^{m'm} (y)\gamma_\mu]\\
\notag
\Pi_{\mu\nu}^{\texttt{QCD}3}(p,q)=&&\frac{1}{\sqrt2}\int d^4xd^4ye^{ip\cdot x}e^{iq \cdot y}Tr[U^{nn'}(-x)\gamma_\nu Q^{n'n}(x)\gamma_\mu] Tr[S^{mm'}(-y)\gamma_5U^{m' m}(y)\gamma_5]\\
\notag
\Pi_{\mu\nu}^{\texttt{QCD}4}(p,q)=&&\frac{1}{\sqrt2}\int d^4xd^4ye^{ip\cdot x}e^{iq \cdot y}Tr[U^{nn'}(-x)\gamma_\mu Q^{n'n}(x)\gamma_5] Tr[S^{mm'}(-y)\gamma_5U^{m'm} (y)\gamma_\nu]\\
\notag
\Pi^{\texttt{QCD}5}(p,q)=&&-\frac{i}{\sqrt3}q_\rho\int d^4xd^4ye^{ip\cdot x}e^{iq \cdot y}Tr[U^{nm'}(-y)\gamma_\rho\gamma_5Q^{m'm}(y)\gamma_5S^{mn'}(-x) \gamma_5U^ {n'n}(x)\gamma_5]\\
\notag
\Pi_{\lambda\sigma}^{\texttt{QCD}6}(p,q)=&&\frac{i}{\sqrt3}q_\rho (g_{\lambda \mu }-\frac{p_\lambda p_\mu}{p^2})(g_{\sigma\nu}-\frac{{p'}_\sigma{p'}_\nu}{{p'}^2}) \int d^4xd^4ye^{ip\cdot x}e^{iq\cdot y}Tr[U^{nm'}(-y)\gamma_\rho\gamma_5Q^{m'm} (y)\\
&&\times\gamma_5S^{mn'}(-x) \gamma_\nu U^{n'n}(x)\gamma_\mu]
\end{eqnarray}
\end{widetext}
where $S^{ij}(x)$, $U^{ij}(x)$ and $Q^{ij}(x)$ are the full propagators of quark field $s$, $u/d$ and $c/b$, respectively. They can be represented as,
\begin{widetext}
\begin{eqnarray}\label{light quark propagator}
\notag
q^{ij}(x)=&&\frac{i\delta^{ij}\slashed{x}}{2\pi^2x^4}-\frac{\delta^{ij}m_q}{2\pi^2x^2}-\frac{\delta^{ij}\left\langle{\bar qq}\right\rangle}{12}+ \frac{i\delta^{ij} \slashed{x}m_q\left\langle{\bar qq}\right\rangle}{48}-\frac{\delta^{ij}x^2\left \langle{\bar qg_s\sigma Gq} \right\rangle}{192}+\frac{i\delta^{ij}x^2\slashed{x}m_q \left\langle{\bar qg_s\sigma Gq}\right\rangle}{1152}-\frac{i\delta^{ij}x^2\slashed {x}g_s^2{\left\langle{\bar qq}\right\rangle}^2}{7776}\\
&&-\frac{ig_sG_{\alpha\beta}^at_{ij}^a(\slashed{x}\sigma^{\alpha\beta}+\sigma^{\alpha\beta}\slashed{x})}{32\pi^2x^2}-\frac{\delta^{ij}x^4\left\langle{\bar qq}\right \rangle\left\langle g_s^2GG\right\rangle}{27648}-\frac{\left\langle{\bar q}^j \sigma^{\mu\nu}q^i\right\rangle\sigma_{\mu\nu}}{8}-\frac{\left\langle{\bar q}^j \gamma^\mu q^i\right\rangle{\gamma_\mu}}{4}+\cdot\cdot\cdot
\end{eqnarray}
\begin{eqnarray}\label{heavy quark propagator}
\notag
Q^{ij}(x)=&&\frac{i}{(2\pi)^4}\int d^4ke^{-ik\cdot x}\left\{\frac{\delta^{ij}( \slashed{k}+m_Q)}{(k^2-m_Q^2)}-\frac{g_sG_{\alpha\beta}^nt_{ij}^n}{4}\frac{\sigma^{\alpha\beta}(\slashed{k}+m_Q)+(\slashed{k}+m_Q)\sigma^{\alpha\beta}}{(k^2-m_Q^2)^2}+ \frac{g_sD_\alpha G_{\beta\lambda}^nt_{ij}^n(f^{\lambda\beta\alpha}+f^{\lambda \alpha\beta})}{3(k^2-m_Q^2)^4}\right.\\
\notag
&&\left.-\frac{g_s^2(t^at^b)_{ij}G_{\alpha\beta}^aG_{\mu\nu}^b(f^{\alpha\beta\mu\nu}+f^{\alpha\mu\beta\nu}+f^{\alpha\mu\nu\beta})}{4(k^2-m_Q^2)^5}+\frac{\delta_{ij}\left\langle{g_s^3GGG}\right\rangle(\slashed k+m_Q)[\slashed k(k^2-3m_Q^2)+2m_Q(2k^2- m_Q^2)](\slashed k+m_Q)}{48(k^2-m_Q^2)^6}+\cdot\cdot\cdot\right\}\\
\end{eqnarray}
\end{widetext}
Here, $q^{ij}(x)$ denotes the light quark propagators $S^{ij}(x)$ and $U^{ij}(x)$. The superscript $i$ and $j$ are color indices, and $t^n=\lambda^n/2$, the $\lambda^n$ is the Gell-Mann matrix, $\sigma_{\alpha\beta} =i[\gamma_\alpha,\gamma _\beta]/2$, $D_\alpha=\partial_\alpha-ig_sG_\alpha^nt^n$ with $G_\alpha^n$ being the gluon field. In the heavy quark propagator $Q^{ij}$, $f^{\lambda\beta\alpha}$ and $f^{\alpha\beta\mu\nu}$ have the following expressions,

\begin{eqnarray}\label{f of double GG definition}
\notag
f^{\lambda\beta\alpha}=&&(\slashed{k}+m_Q)\gamma^\lambda(\slashed{k}+m_Q)\gamma^\beta(\slashed{k}+m_Q)\gamma^\alpha(\slashed{k}+m_Q)\\
\notag
f^{\alpha\beta\mu\nu}=&&(\slashed{k}+m_Q)\gamma^\alpha(\slashed{k}+m_Q)\gamma^\beta(\slashed{k}+m_Q)\\
&&\times \gamma^\mu(\slashed{k}+m_Q)\gamma^\nu(\slashed{k}+m_Q)
\end{eqnarray}
The above full propagators are putted into Eq. (\ref{contraction of TDK}) to calculate the correlation functions in the QCD side. Then, the correlation function in the QCD side can be expanded in different tensor structures,
\begin{eqnarray}\label{scalar invariant amplitude}
\notag
\Pi^{\texttt{QCD}1}(p,q)=&&\Pi_1^{\texttt{QCD}1}\\
\notag
\Pi_{\mu\nu}^{\texttt{QCD}2}(p,q)=&&\Pi_1^{\texttt{QCD}2}g_{\mu\nu}+\Pi_2^{\texttt{QCD}2}p_\mu p_\nu\\
\notag
\Pi_{\mu\nu}^{\texttt{QCD}3}(p,q)=&&\Pi_1^{\texttt{QCD}3}g_{\mu\nu}+\Pi_2^{\texttt{QCD}3}q_\mu q_\nu\\
\notag
\Pi^{\texttt{QCD}5}(p,q)=&&\Pi_1^{\texttt{QCD}5}\\
\notag
\Pi_{\lambda\sigma}^{\texttt{QCD}6}(p,q)=&&\Pi_1^{\texttt{QCD}6}g_{\lambda\sigma}+\Pi_2^{\texttt{QCD}6}p_\lambda q_\sigma\\
&&+\Pi_3^{\texttt{QCD}6}p'_\lambda p'_\sigma+\Pi_4^{\texttt{QCD}6}p'_\lambda q_\sigma
\end{eqnarray}
where $\Pi^{\texttt{QCD}i}_{j}$ are scalar invariant amplitudes, and are composed of perturbative and non-perturbative terms,
\begin{eqnarray}
\Pi_j^\texttt{QCD}=\Pi_j^\texttt{pert}+\Pi_j^\texttt{non-pert}
\end{eqnarray}
where non-perturbative terms include $\left\langle\bar qq\right\rangle$, $\left\langle g_s^2GG\right\rangle$, $\left\langle\bar qg_s\sigma Gq\right\rangle $, $\left\langle g_s^3GGG\right\rangle$ and $\left\langle\bar qq\right\rangle^2$ in the present work. The contributions of higher dimensions than 6 can be safely ignored because they are further suppressed by $O(\alpha_s^2)$.

According to double dispersion relation, the scalar invariant amplitude can be written as,
\begin{eqnarray}
\Pi_j^\texttt{QCD}(p{'^2},p^2,q^2)=\int\limits_{s_t}^{\infty}ds\int\limits_{u_t}^{\infty}du\frac{\rho_j^\texttt{QCD}(p{'^2},s,u)}{(s-p^2)(u-q^2)}
\end{eqnarray}
where $s=p^2$, $u=q^2$ and $p'=p+q$.

Taking strong coupling constant $G_{T_{DK}DK}$ as an example, we will show how QCD spectral densities $\rho_j^\texttt{QCD}$ are obtained. In order to obtain the expression of its perturbation term, the first term of full propagator in Eq. (\ref{heavy quark propagator}) is substituted into Eq. (\ref{contraction of TDK}). After the integration in coordinate and momentum spaces, the perturbative term of correlation function can be written as,

\begin{figure}
	\centering
	\includegraphics[width=7cm, trim=10 90 10 80, clip]{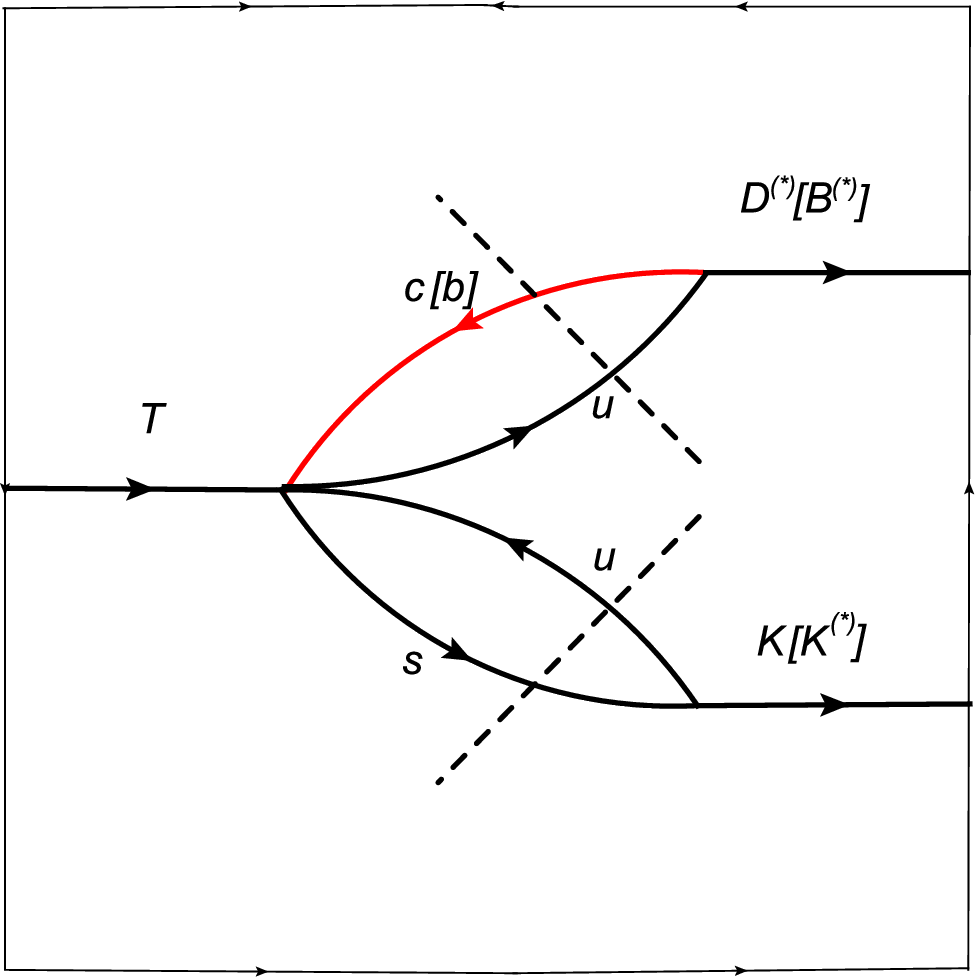}
	\caption{The Feynman diagram for the perturbative part, The dashed lines denote the Cutkosky's cuts.}\label{pertturbative}
\end{figure}

\begin{widetext}

\begin{eqnarray}
\notag
\Pi^{\texttt{QCD}1}(p,q)=&&\frac{-9}{\sqrt2(2\pi)^8}\int d^4k_2\frac{1}{[(k_2-p)^2- m_q^2](k_2^2-m_c^2)}Tr[(\slashed{k}_2-\slashed{p}+m_q)\gamma_5(\slashed{k}_2+m_c)\gamma_5]\\
\notag
&&\times\int d^4k_3\frac{1}{(k_3^2-m_q^2)[(k_3-q)^2-m_s^2]}Tr(\slashed{k}_3+m_q) \gamma_5(\slashed{k}_3-\slashed{q}+m_s)\gamma_5]
\end{eqnarray}

\end{widetext}
According to Cutkosky's rule, we set all quark lines on shell to obtain the QCD spectral density of perturbative term. The corresponding feynman diagram is shown in Fig. \ref{pertturbative}.

\begin{figure*}[h!]
	\centering
	\includegraphics[width=16cm, trim=90 140 60 60, clip]{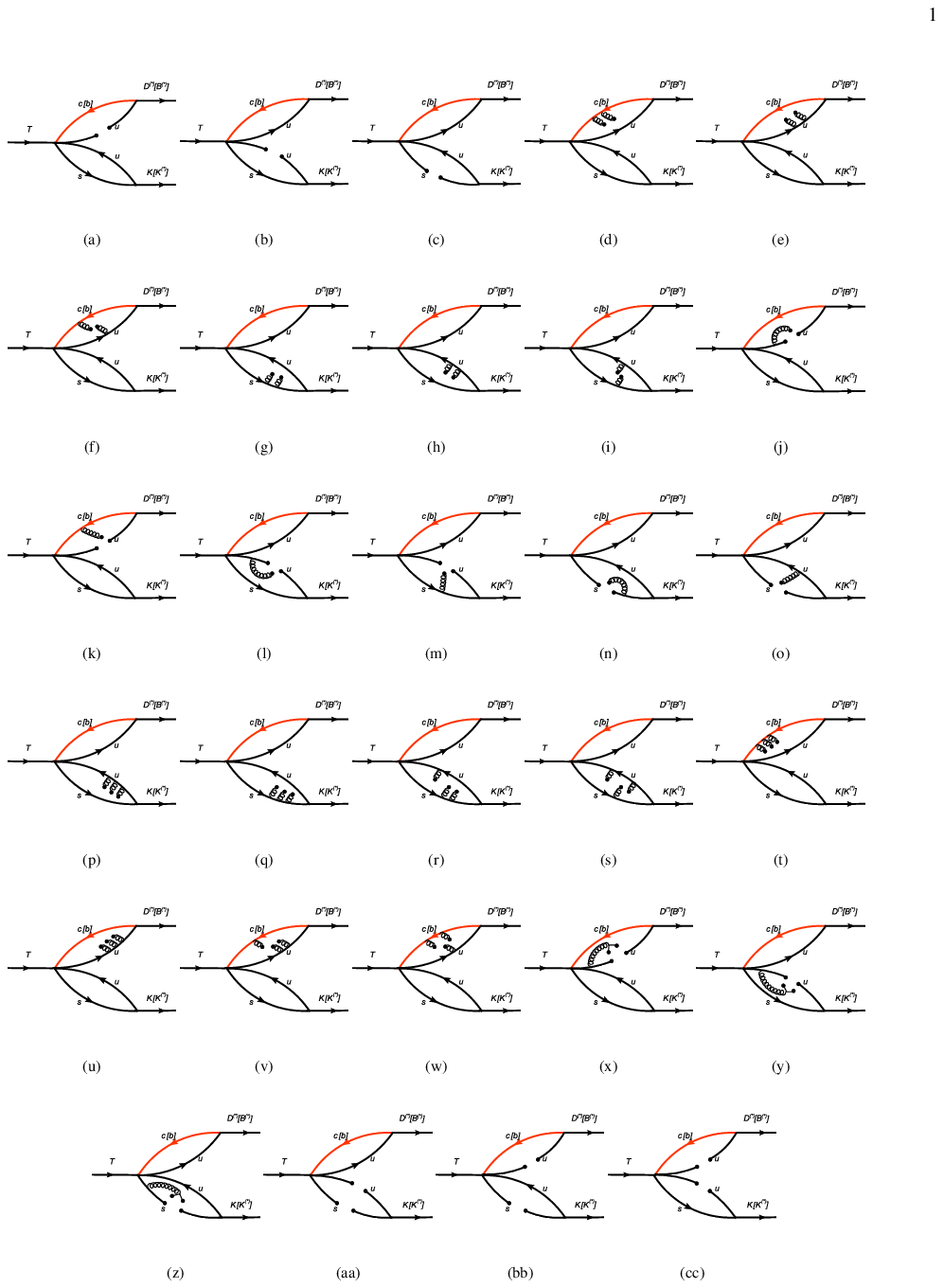}
	\caption{The Feynman diagrams of vacuum condensates for molecule states $D[B]K,D^*[B^*]K,D[B]K^*$, where black and red line represent light and heavy quark respectively, and T denotes the hadronic molecule.}\label{feynman pic}
\end{figure*}

\begin{widetext}
\begin{eqnarray}
\notag
\rho^\texttt{QCD1}_\texttt{pert}=&&\frac{-9}{\sqrt2(2\pi)^8}\int d^4k_2\delta(k_2^2-m_c^2) \delta[(k_2-p)^2-m_q^2]Tr[(\slashed{k}_2-\slashed{p}+m_q)\gamma_5(\slashed{k}_2+m_c)\gamma_5]\\
\notag
&&\times\int d^4k_3\delta(k_3^2-m_q^2)\delta[(k_3-q)^2-m_s^2]Tr(\slashed{k}_3+m_q) \gamma_5(\slashed{k}_3-\slashed{q}+m_s)\gamma_5]\\
\notag
=&&\frac{-9}{\sqrt2(2\pi)^8}\frac{\pi\sqrt{\lambda(s,m_c^2,m_q^2)}}{2s}2(m_c^2-s)\frac{\pi\sqrt{\lambda(u,m_q^2,m_s^2)}}{2u}2(m_s^2-u)
\end{eqnarray}
where $\sqrt{\lambda(a^2,b^2,c^2)}=\sqrt{(a^2+b^2-c^2)^2-4a^2b^2}$.
\end{widetext}
The calculation process of vacuum condensates which is shown in Fig. \ref{feynman pic} is similar to that of perturbative part, except for the following integration,

\begin{eqnarray}
I_{ab}=\int d^4k\frac{1}{(k^2-m_1^2)^a[(k-p)^2-m_2^2]^b}
\end{eqnarray}
By taking the derivative and reducing the power of denominators, this integration can still be processed by Cutkosky's rules,
\begin{eqnarray}
\notag
I_{ab}=&&\frac{\partial_{m_1^2}^{a-1}\partial_{m_2^2}^{b-1}}{(a-1)!(b-1)!}\int {d^4 k}\frac{1}{(k^2-m_1^2)[(k-p)^2-m_2^2]}\\
&&\to\frac{(-2\pi i)^2}{2\pi i}\frac{\partial_{m_1^2}^{a-1} \partial _{m_2^2}^{b-1}}{(a-1)!(b-1)!}\frac{\pi\sqrt{\lambda(s,m_1^2,m_2^2)}}{2s}
\end{eqnarray}

After obtaining the spectral density, we match the phenomenological side and QCD side according to the quark-hadron duality. In method I, the momentum $p^{\prime}$ is set to be $p^{\prime2}=kp^2$. While in method II, the initial hadron is set to be off-shell with $P'^2=-p'^2>0$. Then, we perform double Borel transformation with regard to the variables $P^2=-p^2$ and $Q^2=-q^2$, as it can enhance the contribution of ground state and suppress that from excited and continuum states. The variables $P^2$ and $Q^2$ will be replaced by Borel parameters $T_1^2$ and $T_2^2$. Furthermore, we also introduce $T_2^{2}=k'T_1^{2}=k'T^2$ to simplify the calculation. Finally, we obtain the QCD sum rules of these coupling constants.

\section{Numerical result of strong coupling constants}\label{sec3}
The final numerical results are dependent on some input parameters such as the masses of heavy quarks and vacuum condensates. It is noted that the the masses of $c$ and $b$ quarks and the values of vacuum condensates are energy scale dependent, which can be expressed as follows by using the renormlization group equation (RGE),
\begin{eqnarray}\label{vacuum codensation}
\notag
&&m_Q(\mu)=m_Q(m_Q)\left[{\frac{\alpha_s(\mu)}{\alpha_s(m_Q)}} \right]^{\frac{12} {33-2n_f}}\\
\notag
&&m_s(\mu)=m_s(2\mathrm{GeV})\left[{\frac{\alpha_s(\mu)}{\alpha_s(2\mathrm{GeV})}} \right]^{\frac{12}{33-2n_f}}\\
\notag
&&\left\langle\bar qq\right\rangle(\mu)=\left\langle\bar qq\right\rangle(\mathrm{1Ge V}){\left[{\frac{\alpha_s(\mathrm{1GeV})}{\alpha_s(\mu)}} \right]^{\frac{12}{33-2n _f}}}\\
\notag
&&\left\langle\bar qg_s\sigma Gq\right\rangle(\mu)=\left\langle\bar qg_s\sigma Gq \right\rangle(\mathrm{1GeV})\left[{\frac{\alpha_s(\mathrm{1GeV})}{\alpha_s(\mu)}} \right]^{\frac{2}{33-2n_f}}\\ \notag
&&\alpha_s(\mu)=\frac{1}{b_0t}\left[1-\frac{b_1}{b_0^2}\frac{\log t}{t}+\frac{b_1^2 (\log^2t-\log t-1)+b_0b_2}{b_0^4t^2}\right]\\
\end{eqnarray}
where $q$ = $u$, $d$ and $s$ for the condensate term, $t=\log \frac{\mu^2}{\Lambda_{QCD}^2},$ $b_0=\frac{33-2n_f}{12\pi},$ $b_1=\frac{153-19n_f}{24\pi^2},$ $b_2=\frac{2857-\frac{5033}{9}n_f+\frac{325}{27}n_f^2}{128\pi^3}$, and $\Lambda_{QCD}$ = 210, 292, 332 MeV for the flavors $n_f$ = 5, 4, 3 \cite{ParticleDataGroup:2024cfk}, respectively. In the present work, flavor numbers are set to be $n_f$ = 4 and 5 for the charm-strange and bottom-strange tetraquark molecular states.
	
In our previous work, the energy scale $\mu = 1$, $1.1$, and $2.5$ GeV for $D^{(*)}, D_s^{(*)}$, and $B^{(*)}$ works well in analyzing the properties of these particles \cite{Wang:2015mxa}, so these values are still employed in this work. The standard values of vacuum condensates are taken as $\left\langle \bar qq \right\rangle =-(0.24\pm0.01\ \mathrm{GeV})^3,$ $\left\langle\bar ss\right\rangle=(0.8\pm0.1)\left\langle \bar qq\right\rangle,$ $\left\langle\bar qg_s\sigma Gq \right\rangle=m_0^2\left\langle\bar qq\right\rangle,$ $\left\langle\bar sg_s\sigma Gs\right\rangle=m_0^2\left\langle\bar ss\right\rangle,$ $m_0^2=(0.8\pm0.1)\ \mathrm{GeV^2},$ $\left\langle{\frac{\alpha_sGG}{\pi }}\right\rangle=(0.012\pm0.004)\ \mathrm{GeV^4}$ at the energy scale $\mu$=1 GeV with $q=u$ and $d$ quarks \cite{Shifman:1978by, Shifman:1978bx, Reinders:1984sr}. The modified-minimal-subtraction masses are adopted as $m_c(m_c)=(1.275\pm0.025),$ GeV $m_b(m_b)=(4.18\pm0.03)\ \mathrm{GeV}$ and $m_s(2\mathrm{GeV})=(0.095\pm0.005)\ \mathrm{GeV}$ from the Particle Data Group \cite{ParticleDataGroup:2024cfk}. In addition, we adopt $\frac{f_Km_K^2}{m_s+m_u}=-\frac{\left\langle\bar qq\right\rangle+\left\langle \bar ss\right\rangle}{f_K(1-\delta_K)}$ from Gell-Mann-Oakes-Renner relation with $\delta_K=0.50$ \cite{Bordes:2012ud}. The other input parameters are listed in Table \ref{input parameters of three point correlator}.
\begin{table*}[htbp]
	\begin{ruledtabular}
		\renewcommand{\arraystretch}{1.4}
	\caption{Input parameters used in calculation of three-point correlation function \cite{Zhou:2025yjb,Wang:2019iaa,Wang:2022ckc,Wang:2013ff,Wang:2011fv}.}
		\begin{tabular}{c >{\centering}p{7em} c c c c c c c}
				Parameters & Values (GeV) & Parameters & Values (GeV) & Parameters & Values (GeV) & Parameters & Values\\
				\hline
				$m_D$ & 1.864 &$f_D$ & 0.208 & $m_K$ & 0.493 & $f_K$ & 0.156 GeV\\
				$m_B$ & 5.28 &$f_B$ & 0.189 & $m_{B^*}$ & 5.324 &$f_{T_{BK^*}}$ & $2.60 \times 10^{-2}$ GeV$^5$\\
				$m_{D^*}$ & 2.007 &$f_{D^*}$ & 0.263 & $m_{K^*}$ & 0.892 & $f_{T_{BK}}$ & $1.77 \times 10^{-2}$ GeV$^5$\\
				$m_{D_s}$ & 1.969 &$f_{D^*}$ & 0.240 & $m_{D_s^*}$ & 2.112 &$f_{T_{D^*K}}$ & $3.88 \times 10^{-3}$ GeV$^5$\\
				$m_\eta$ & 0.5478 &$f_\eta$ & 0.145 & $m_{T_{DK}}$ & 2.322 & $f_{T_{DK}}$ & $3.20 \times 10^{-3}$ GeV$^5$\\
				$m_{T_{D^*K}}$ & 2.457 & $f_{D_s^*}$ & 0.308 & $m_{T_{DK^*}}$ & 2.538 & $f_{T_{DK^*}}$ & $4.89 \times 10^{-3}$ GeV$^5$\\
				$m_{T_{BK}}$ & 5.970 & $f_{K^*}$ & 0.220 & $m_{T_{B^*K}}$ & 6.050 & $f_{T_{B^*K}}$ & $2.05 \times 10^{-2}$ GeV$^5$\\
				$m_{T_{BK^*}}$ & 6.158 & $f_{B^*}$ & 0.195 & $\sqrt {s_D^0} $ & 2.49 & $\sqrt {u_K^0}$ & 1 GeV\\
				$\sqrt {s_{D^*}^0} $ & 2.5 &$\sqrt {u_{K^*}^0} $ & 1.3 & $\sqrt {s_{D_s}^0} $ & 2.5 & $\sqrt {s_{D_s^*}^0}$ & 2.6 GeV\\
				$\sqrt {s_\eta^0} $ & 0.9478 & $\sqrt {s_B^0} $ & $5.79$   & $\sqrt {s_{B^*}^0} $  & $5.9$ & $-$ & $-$
		\end{tabular}
		\label{input parameters of three point correlator}
	\end{ruledtabular}
\end{table*}

The sum rules of coupling constants are dependent on Borel parameter $T^2$. To obtain the reliable results, an appropriate work region should be selected, where the coupling constants have a weak Borel parameter dependency. Furthermore, the convergence of OPE should be satisfied. The contributions of the vacuum condensates of dimension $n$ can be expressed as,
\begin{eqnarray}
\mathrm{D(n)}=\frac{\int\limits_{s_t}^{s_0}ds\int\limits_{u_t}^{u_0}du\rho_n^\texttt{QCD}(s,u)\exp\left(-\frac{s}{T_1^2}-\frac{u}{T_2^2}\right)}{\int\limits_{s_t}^{s_0}ds\int\limits_{u_t}^{u_0}du\rho^\texttt{QCD}(s,u)\exp\left(-\frac{s}{T_1^2}-\frac{u}{T_2^2}\right)}
\end{eqnarray}
where $\rho_n^\texttt{QCD}$ is spectral density of $n$ dimension vacuum condensate.

\begin{figure}[htbp]
	\centering
	\includegraphics[width=8.5cm, trim=90 30 60 30, clip]{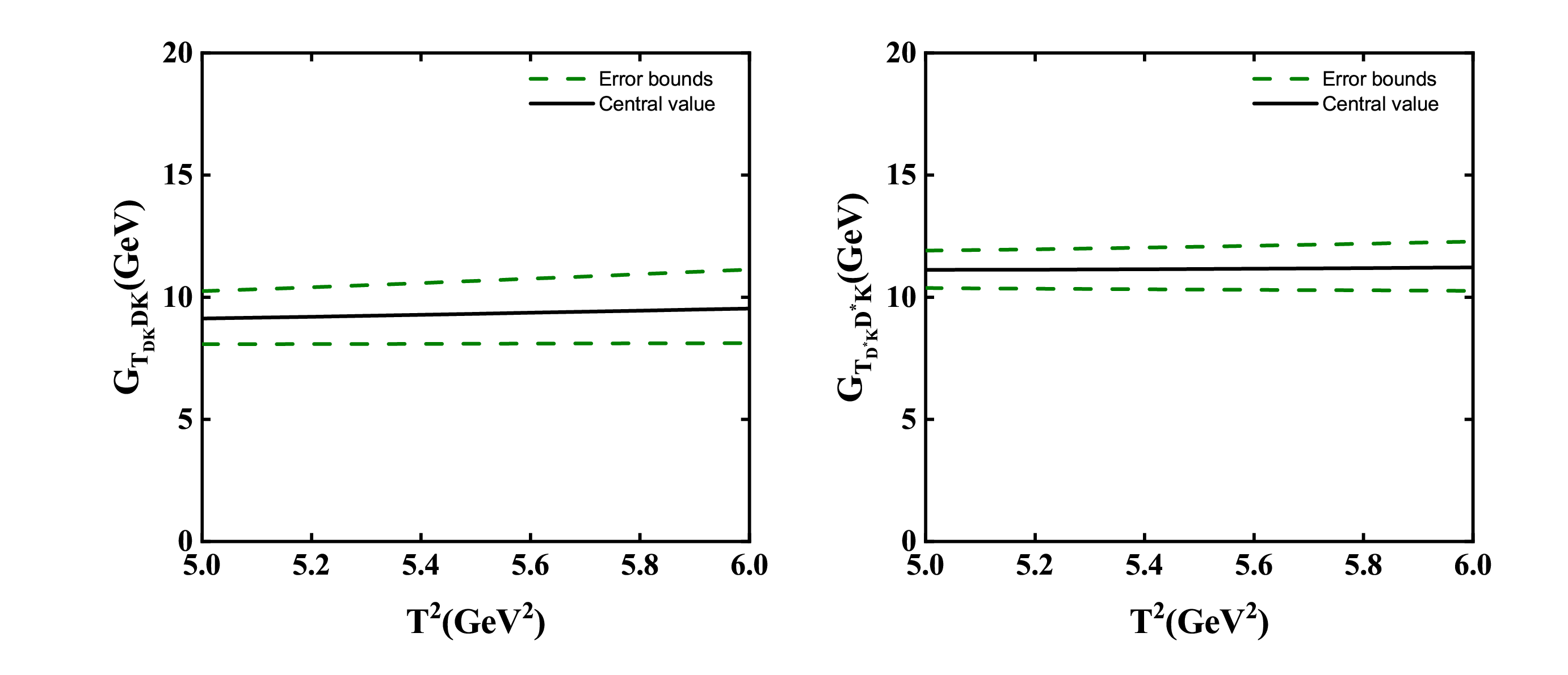}
	\caption{The central values of hadronic coupling constants $G_{T_{DK}DK}$ and $G_{T_{D^*K}D^*K}$ with variations of the Borel parameters $T^2$ in the Borel platform.}
	\label{1}
	\centering
	\includegraphics[width=8.5cm, trim=90 30 60 30, clip]{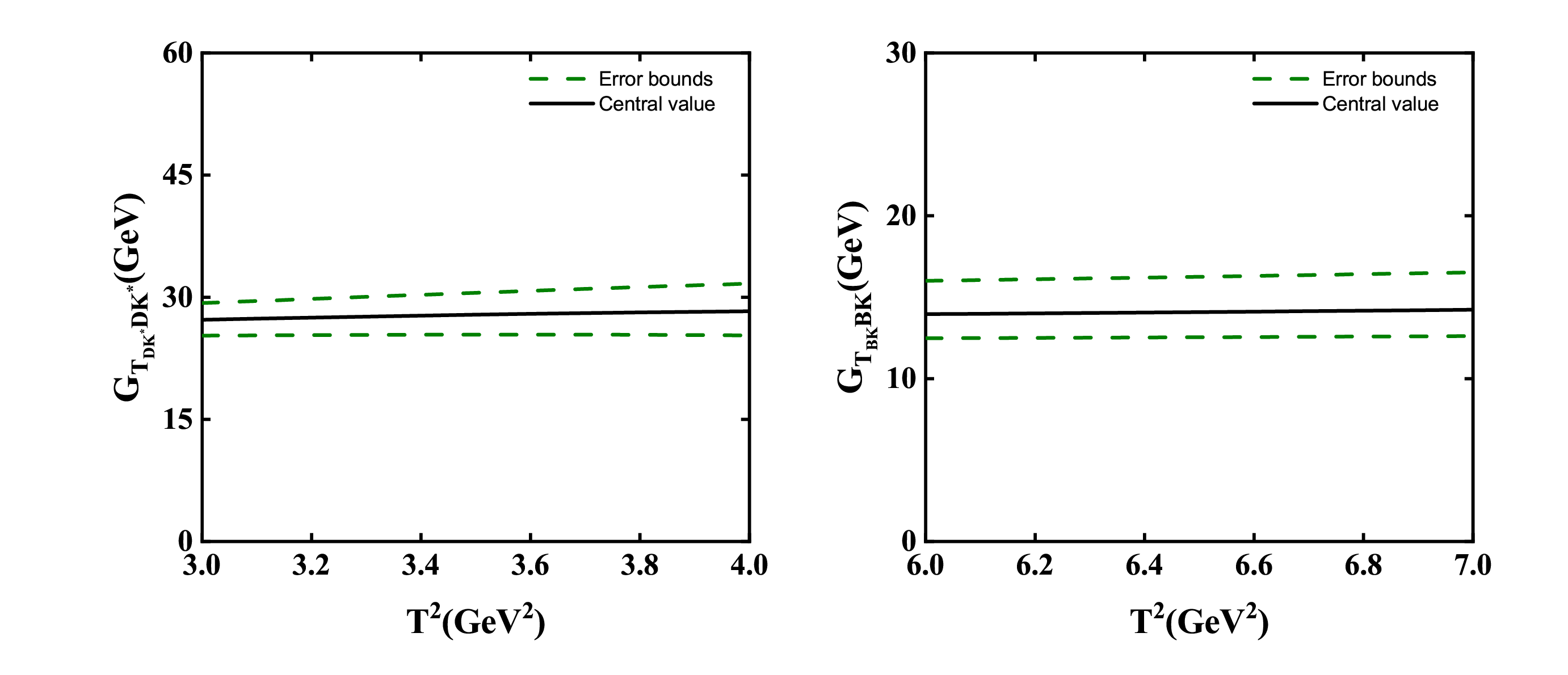}
	\caption{Same as Fig. 4 but for $G_{T_{DK^*}DK^*}$ and $G_{T_{BK}BK}$.}
	\label{2}
	\centering
	\includegraphics[width=8.5cm, trim=90 30 60 30, clip]{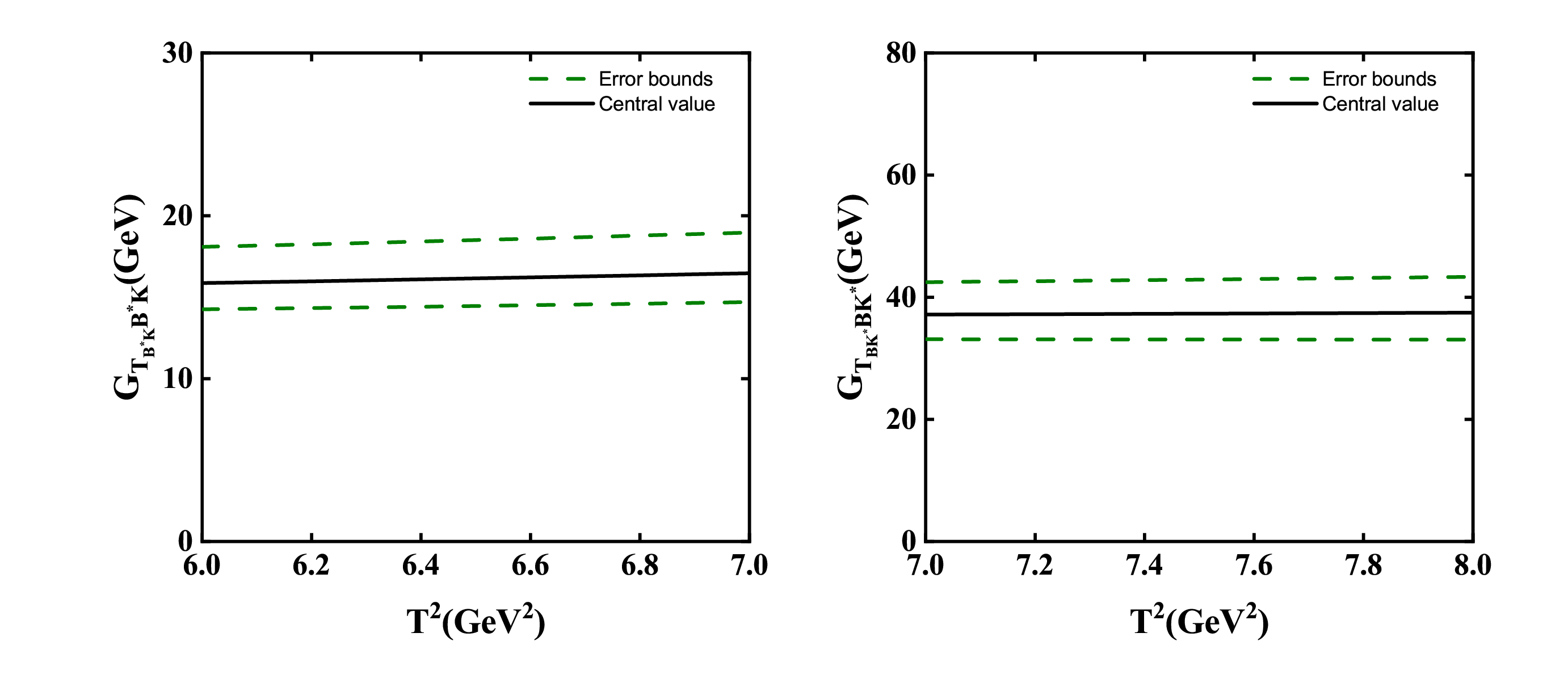}
	\caption{Same as Fig. 4 but for $G_{T_{B^*K}B^*K}$ and $G_{T_{BK^*}BK^*}$.}
	\label{3}
	\centering
	\includegraphics[width=8.5cm, trim=90 30 60 30, clip]{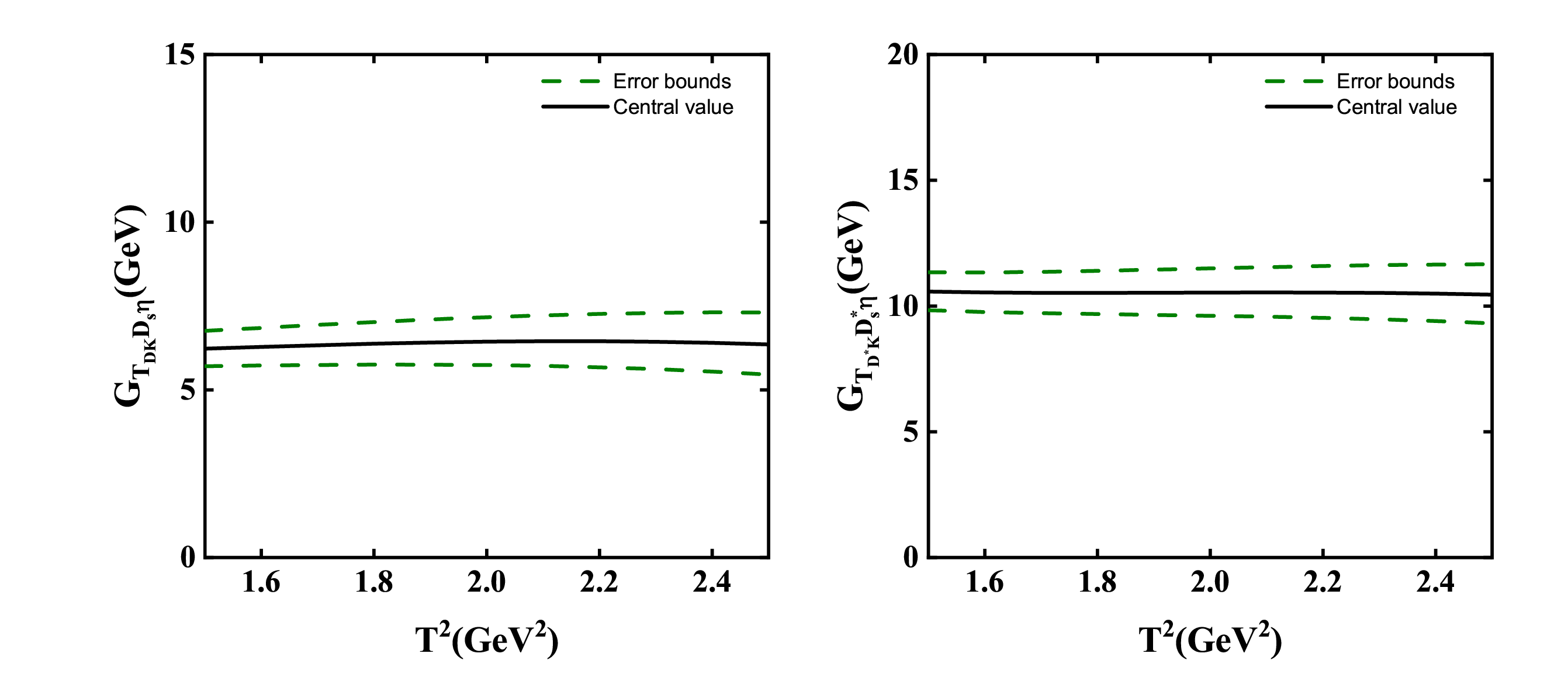}
	\caption{Same as Fig. 4 but for $G_{T_{DK}D_s\eta}$ and $G_{T_{D^*K}D_s^*\eta }$.}
	\label{4}
\end{figure}

In method I, the value of $k$ is set to be $k=1$. As for the value of $k^{\prime}$, it is chosen to be $k^{\prime}=m_2^2/m_1^2$ for charmed section except for coupling constants $G_{T_{D^{(*)}K}D_s^{(*)}\eta}$, while $k'=1$ is selected for bottom section. Then, we adjust suitable $C_i$ to obtain good Borel platforms to extract coupling constants. The values of these parameters and Borel platforms are shown in Table \ref{fitting and cutting}. The Borel platforms of coupling constants are also explicitly illustrated in Figs. \ref{1} $\sim$ \ref{4}.
\begin{table*}
	\centering
	\begin{ruledtabular}
		\centering
		\renewcommand{\arraystretch}{1.4}
		\caption{The results of coupling constants obtained by two methods.}
		\begin{tabular}{c c c c c c}
			Coupling constants & Method I (GeV) & Borel platform $T^2$(GeV$^2$) & $C_i$($\times10^{-4}$ ) & Method II (GeV) &  Borel platform $T^2$ (GeV$^2$)\\
			\hline
			$G_{T_{DK}DK}$ & $9.32_{-1.25}^{+1.81}$ & $5-6$ & $1.3T^2$ GeV$^6$ & $10.40_{-0.65}^{+0.70}$ & $4-5$ \\
			$G_{T_{D^*K}D^*K}$ & $11.15_{-0.89}^{+1.12}$ & $5-6$ & $-1.4T^2$ GeV$^6$+$5.18$ GeV$^8$ & $8.53_{-0.47}^{+0.50}$ & $4-5$ \\
			$G_{T_{DK^*}DK^*}$ & $27.84_{-2.57}^{+3.83}$ & $3-4$ & $-6T^2$ GeV$^6$+$12$ GeV $^8$ & $32.77_{-2.53}^{+3.10}$ & $2-4$ \\
			$G_{T_{BK}BK}$ & $14.08_{-1.60}^{+2.44}$ & $6-7$ & $5T^2$ GeV$^6$-$20.5$ GeV$^8$& $12.07_{-1.18}^{+1.72}$ & $7-9$ \\
			$G_{T_{B^*K}B^*K}$ & $16.15_{-1.90}^{+2.82}$ & $6-7$ & $-4T^2$ GeV$^6$+$17.6$ GeV$^8$& $13.32_{-1.29}^{+1.88}$ & $7-9$ \\
			$G_{T_{BK^*}BK^*}$ & $37.29_{-4.24}^{+6.04}$ & $6-7$ & $-8T^2$ GeV$^6$& $36.09_{-2.86}^{+3.77}$ & $6-8$\\
			$G_{T_{DK}D_s\eta}$ & $6.43_{-0.98}^{+0.87}$ & $1.5-2.5$ & $-0.98T^2$ GeV$^7$& $-$ & $-$ \\
			$G_{T_{D^*K}D_s^*\eta}$ & $10.53_{-1.22}^{+1.12}$ & $1.5-2.5$ & $-1.2T^2$ GeV$^7$& $-$ & $-$
		\end{tabular}
		\label{fitting and cutting}
	\end{ruledtabular}
\end{table*}
It can be seen from these figures that the final results are stable and reliable on the Borel parameters $T^{2}$. In addition, the contributions of perturbative part and different condensate terms are illustrated in Fig. \ref{contribution of dimension} in the Appendix A. It is shown by these figures that the contribution from higher condensate term such as $D(6)$ is less than $5\%$, which means that the condition of convergence of OPE is satisfied. As for the uncertainty of results, it mainly comes from input parameters in the QCD side, and can be absorbed into decay constants (or pole residues) and coupling constants. Taking coupling constant $G_{T_{DK}DK}$ as an example, if we use the notation $\alpha$ to stand for the input parameters, the uncertainty $\alpha\to\alpha+\delta\alpha$ will eventuate the uncertainties $f_Df_KG_1f_{T_{DK}}\to\bar f_D\bar f_K\bar G_1\bar f_{T_{DK}}+ \delta f_Df_KG_1f_{T_{DK}}$, $C_1\to\bar C_1+\delta{C_1}$ with
\begin{eqnarray}
\delta f_Df_KG_1f_{T_{DK}}=\bar f_D\bar f_K\bar G_1\bar f_{T_{DK}}(\frac{\delta f_D}{\bar f_D}+\frac{\delta f_K}{\bar f_K}+\frac{\delta G_1}{\bar G_1}+\frac{\delta f_{T_{DK}}}{\bar f_{T_{DK}}})
\end{eqnarray}
where the overline represents the central value of parameters. To avoid overestimating the uncertainties of coupling constants, we set $\frac{\delta f_D}{\bar f_D}= \frac{\delta f_K}{\bar f_K}=\frac{\delta G_1}{\bar G_1}=\frac{\delta f_{T_{DK}}}{\bar f_{T_{DK}}}$,
\begin{eqnarray}\label{uncertainty}
\delta f_Df_KG_1f_{T_{DK}}=\bar f_D\bar f_K\bar G_1\bar f_{T_{DK}}\frac{4\delta G_1}{\bar G_1}.
\end{eqnarray}
In the present work, we ignore the values of $\delta{C_i}$, and uniformly use Eq. (\ref{uncertainty}) to acquire the uncertainties of coupling constants. According to these analyses, we can obtain reliable results of the strong coupling constants which are displayed in Table \ref{fitting and cutting}.

In method II, the coupling constants are also obtained with $k'$ being consistent with that in method I. Fixing $P'^2=3$ GeV$^2$, the Borel platforms (see Table \ref{fitting and cutting}) are determined after repeated trial and error. By taking different values of $P'^2$ in space-like regions ($P^{\prime2}>0$), the momentum dependent coupling constants $G(P^{\prime2})$ are calculated. In order to obtain its on-shell value $G(P'^2=-m_T^2)$, it is necessary to extrapolate these values into time-like regions ($P^{\prime2}<0$). It is noted that there are no definite expressions for the dependence of coupling constants on $P'^2$. In the present work, the momentum dependent coupling constants can be well fitted by the following exponential function,
\begin{figure}
	\centering
	\includegraphics[width=8.5cm, trim=50 30 170 60,clip]{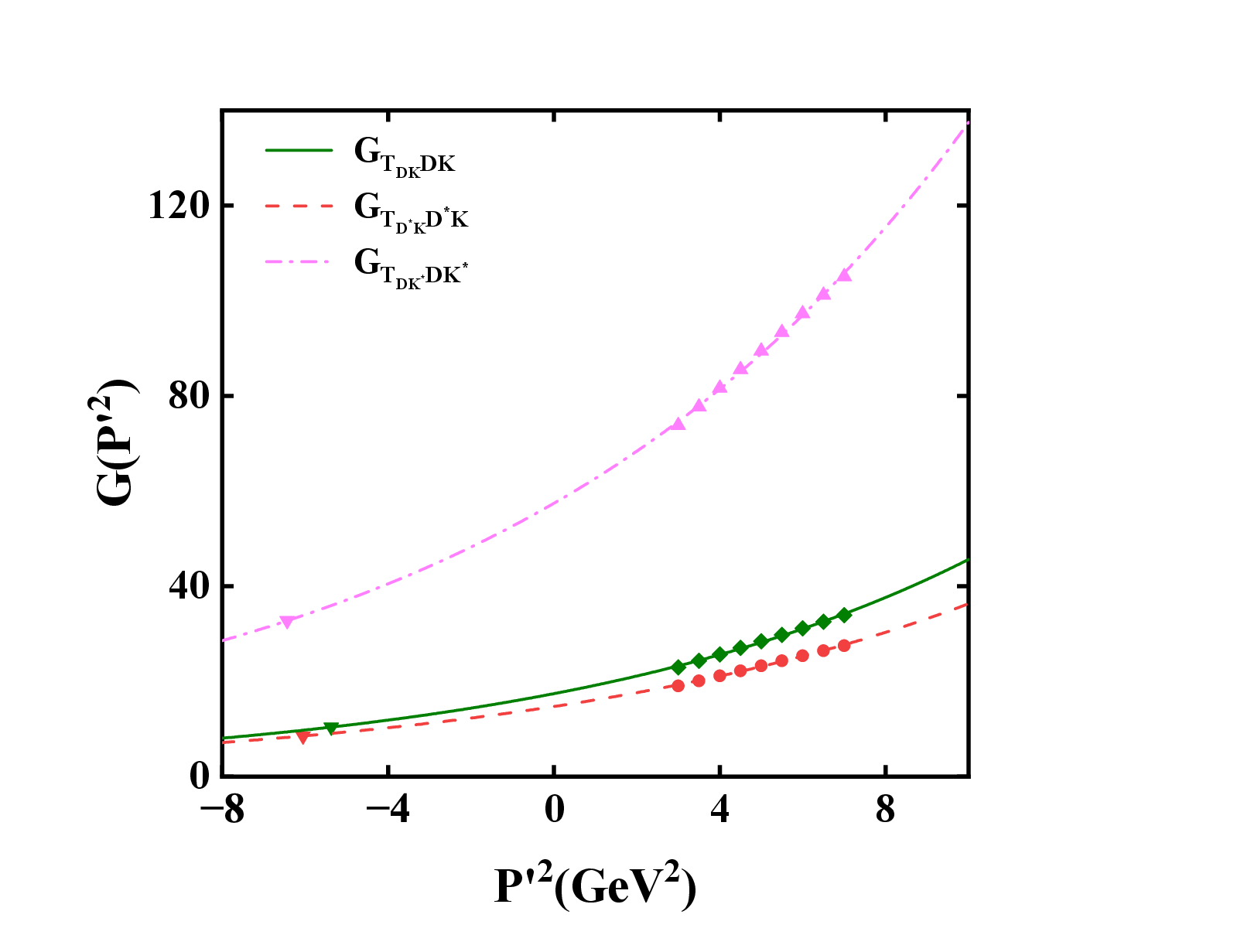}
	\caption{The coupling constants $G_{T_{DK}DK}$, $G_{T_{D^*K}D^*K}$ and $G_{T_{DK^*}DK^*}$ with the variations of ${P'}^2$.}
	\label{couplong constants g1 g2 g3}
	\centering
	\includegraphics[width=8.5cm, trim=50 30 160 50,clip]{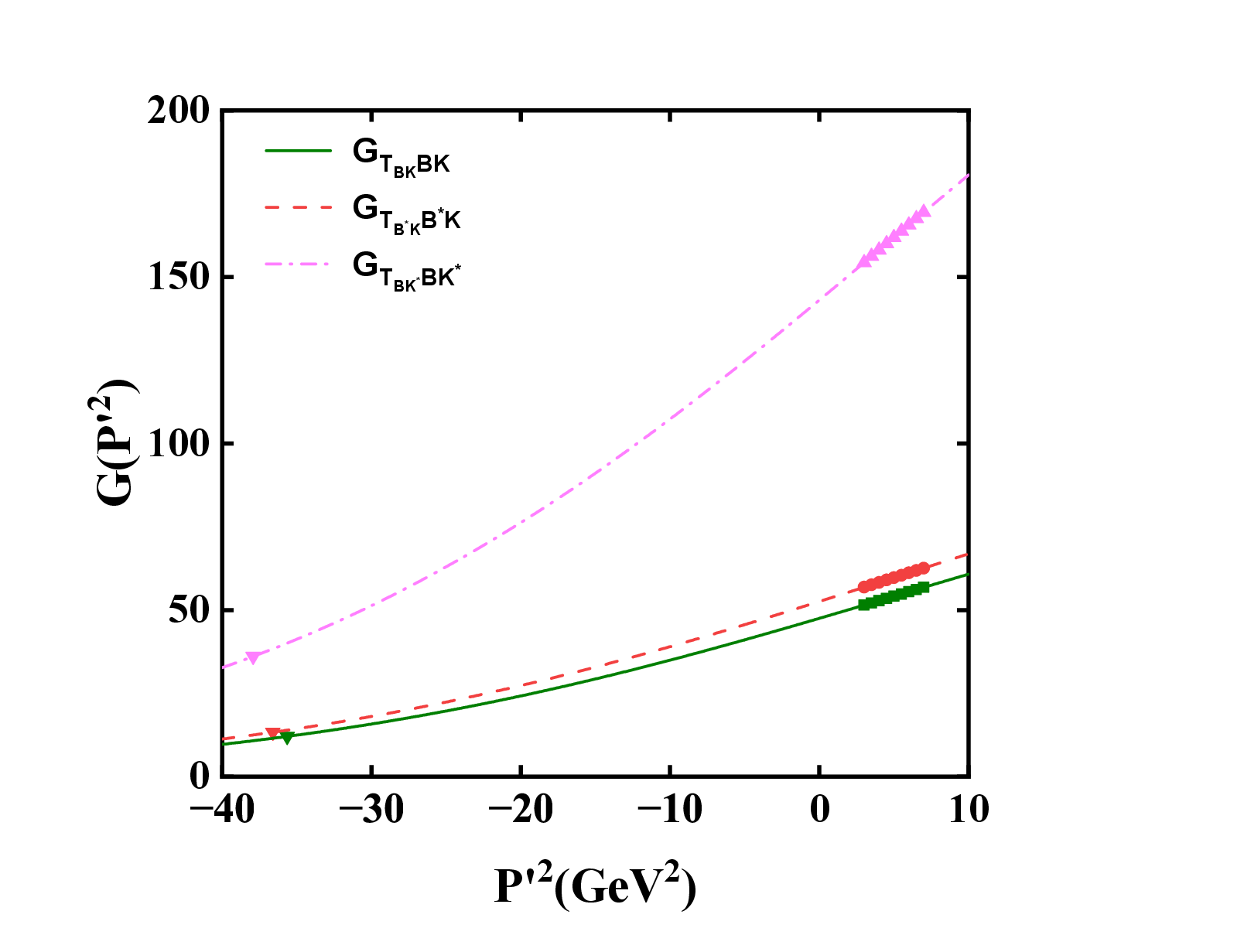}
	\caption{The coupling constants $G_{T_{BK}BK}$, $G_{T_{B^*K}B^*K}$ and $G_{T_{BK^*}BK^*}$ with the variations of ${P'}^2$.}
	\label{couplong constants g4 g5 g6}
\end{figure}
\begin{eqnarray}
G(P{'^2})=a\texttt{exp}[bP^{\prime2}-cP^{\prime4}],
\end{eqnarray}
Here $a$, $b$ and $c$ are fitting parameters, and their values are listed in Table \ref{The values of fitting parameters}. The fitting curves of coupling constants are explicitly shown in Figs. \ref{couplong constants g1 g2 g3} and \ref{couplong constants g4 g5 g6}. From these figures, we can see that the numerical results are well fitted by these functions, and we can obtain reliable on-shell values of the strong coupling constants. By setting the on-shell condition ($P'^2=-m_T^2$) in the fitting function, the strong coupling constants are obtained. The final results are also listed in Table \ref{fitting and cutting}.
\begin{table}
	\centering
	\begin{ruledtabular}
		\centering
		\renewcommand{\arraystretch}{1.4}
		\caption{The values of fitting parameters.}
		\begin{tabular}{c >{\centering}p{7em} c c c}
			Coupling constants & $a$ (GeV) & $b$ ($10^{-2}$GeV$^{-2}$) & $c$ ($10^{-4}$GeV$^{-4}$)\\
			\hline
			$G_{T_{DK}DK}$ & 17.44 & 9.617 & $0$\\
			$G_{T_{D^*K}D^*K}$ & 14.73 & 9.026 & $0$\\
			$G_{T_{DK^*}DK^*}$ & 57.44 & 8.728 & $0$\\
			$G_{T_{BK}BK}$ & 47.51 & 2.765 & 3.027\\
			$G_{T_{B^*K}B^*K}$ & 52.57 & 2.694 & 2.889\\
			$G_{T_{BK^*}BK^*}$ & 143.1 & 2.603 & 2.716
		\end{tabular}
		\label{The values of fitting parameters}
	\end{ruledtabular}
\end{table}

From Table \ref{fitting and cutting}, we can find that the predictions of these two methods are roughly compatible with each other except for $G_{T_{DK}D_s\eta}$ and $G_{T_{D^{*}K}D_s^{*}\eta}$. These consistency indicates the reliability of the final results. It is also noted that we do not obtain reliable values for $G_{T_{DK}D_s\eta}$ and $G_{T_{D^{*}K}D_s^{*}\eta}$ within the framework of method II. It is because we can not find appropriate Borel platform to extract reliable values of the coupling constants. For the sum rules of these two coupling constants, we find that the contribution of dimension 5 is much greater than those of other dimensions, and even accounts for nearly 100 $\%$. The similar situation also occurred in our previous work \cite{Wang:2024qqa}. It has been indicated in Sec. \ref{The phenomenological side} that the excited and continuum
states of initial particles are considered in method I, while not in method II. This above phenomenon may be due to the interference of the excited states of the initial particles.
In Table \ref{fitting and cutting}, we do not list the values of $G_{T_{D[B]K^*}D^*[B^*]K}$ because they are obtained to be zero in both two methods. The reason is that the trace of the correlation function in the QCD side is zero. It means that these interactions are inhibited at the quark level. The decay process $T_{DK^*}\to D^*K$ may occur through the exchange of intermediate particles, and should have a narrow width.

\section{The decay processes of $T_{D^{(*)}K}\to D_{s}^{(*)}\pi$ and $T_{B^{(*)}K}\to B^{(*)}K$}\label{sec4}
As an application of these strong coupling constants, we analyzed the decay processes of $T_{D^{(*)}K}\to D_{s}^{(*)}\pi$ and $T_{B^{(*)}K}\to B^{(*)}K$ with the above predicted results, where the former decay process can be explained by $\eta-\pi$ mixing mechanism. The two-body decay width can be expressed as,
\begin{eqnarray}\label{width}
\Gamma=\frac{1}{2J+1}\sum\frac{p}{8\pi M_i^2}|T|^2
\end{eqnarray}
with $p=\frac{\sqrt{[M_i^2-(m_{f1}+m_{f2})^2][M_i^2-(m_{f1}-m_{f2})^2]}}{2M_i}$. Here, $M_{i}$, and $m_{f1}/m_{f2}$ represent the masses of initial and final particles, $J$ is the total angular momentum of the initial hadron, $\sum$ denotes the summation of all the polarization vectors, and $T$ is the scattering amplitude. With Eqs. (\ref{matrix element}) and (\ref{width}), the decay widths can be written as,
\begin{eqnarray}
\notag
\Gamma_{T_{DK}D_s\pi}=&&t_{\eta\pi}^2G_{T_{DK}D_s\eta}^2\frac{\sqrt{\lambda{(m_{T_{DK}}^2,m_{D_s}^2,m_\pi^2)}}}{16\pi m_{T_{DK}}^3(m_ \eta^2-m_\pi^2)^2}\\
\notag
\Gamma_{T_{D^*K}D_s^*\pi}=&&t_{\eta\pi}^2G_{T_{D^*K} D_s^*\eta}^2\frac{\sqrt{\lambda{(m_{T_{D^*K}}^2,m_{D_s^*}^2,m_\pi^2)}}}{192\pi m_{T_{D^*K}}^5m_{D_s^*}^2(m_\eta^2-m_\pi^2)^2}\\
\notag
&&\times [8m_{T_{D^*K}}^2m_{D_s^*}^2+(m_{T_{D^*K}}^2+m_{D_s^ *}^2-m_\eta^2)^2]\\
\notag
\Gamma_{T_{BK}BK}=&&G_{T_{BK}BK}^2\frac{\sqrt{\lambda{(m_{T_{BK}}^2,m_B^2,m_K^2)}}}{16\pi m_{T_ {BK}}^3}\\
\notag
\Gamma_{T_{B^*K}B^*K}=&&G_{T_{B^*K}B^*K}^2\frac{\sqrt{\lambda{(m_{T_{B^*K}}^2,m_{B^*}^2,m_K^2)}}}{{48\pi m_{T_{B^*K}} ^3}}\\
&&\times [\frac{8m_{T_{B^*K}}^2m_{B^*}^2+(m_{T_{B^*K}}^2+m_{B^*}^2-m_K^2)^2} {4m_{T _{B^*K}}^2m_{B^*}^2}]
\end{eqnarray}
Using the predicted values of the strong coupling constants of method I, we obtain the widths of these above decay channels,
\begin{eqnarray}
\notag
&&\Gamma_{T_{DK}D_s\pi}=10.3_{-2.9}^{+3.0}\; \mathrm{KeV}\\
\notag
&&\Gamma_{T_{D^*K}D_s^*\pi}=24.8_{-5.4}^{+5.5} \; \mathrm{KeV}\\
\notag
&&\Gamma_{T_{BK}BK}=101_{-21}^{+36}\;\mathrm{MeV}\\
&&\Gamma_{T_{B^*K}B^*K}=142_{-25}^{+52}\;\mathrm{MeV}
\end{eqnarray}

To make a comparison, the experimental values and theoretical predictions for $D_{s0}^{*}\to {D_s}\pi^0 $, ${D_{s1}}\to D_s^*\pi^0 $ are all listed in Table \ref{decay width}. Some theoretical results are obtained by treating $D_{s0}^{*}(2317)$ and $D_{s1}(2460)$ as traditional mesons, the others are calculated by supporting them as hadronic molecules or tetraquarks. Form this table, we can see that our results are lower than those in Refs. \cite{Faessler:2007gv,Fu:2021wde,Yue:2025wcl} where the effective Lagrangian approach was employed. And in these literatures, the contributions of $\eta-{\pi^0}$ mixing and meson loops were all considered. If only the $\eta-{\pi^0}$ mixing was included in the calculation, the results are $20\pm2$ and $20\pm3$ KeV for $DK$ and $D^{*}K$ molecular states \cite{Fu:2021wde}. These values are compatible with the results in the present work. If treating them as $c\overline{s}$ mesons, the predicted widths by different groups or with different models are not consistent well with each other, the values range from a few KeV to about two hundred KeV. Whether treated them as $c\overline{s}$ mesons or hadronic molecules, their predicted decaying widths are all in the range of experimental data. It is difficult for us to reach a definite conclusion about the inner-structure of these two particles only according to decay properties.

It is noted that the theoretical masses of $DK$ and $D^{*}K$ hadronic molecules are also analyzed by our previous work, where their predicted masses are consistent with experimental values of $D_{s0}^{*}(2317)$ and $D_{s1}(2460)$. Thus, combining the theoretical analyses about their masses and decay widths, we support the explanation of $D_{s0}^{*}(2317)$ and $D_{s1}(2460)$ as $DK$ and $D^{*}K$ hadronic molecules.
 \begin{table}
	\begin{ruledtabular}
		\renewcommand{\arraystretch}{1.4}
		\caption{The decay widths of $D_{s1}(2460)\to D_s^*\pi^0$ and $D_{s0}^*(2317)\to D_s \pi^0$.}
		\begin{tabular}{c >{\centering}p{7em} c c c}
			Approach & model & $\Gamma (D_{s1}\to D_s^{*}{\pi^0})$ & $\Gamma (D_{s0}^{*} \to D_s{\pi^0})$\\
			\hline
			Ref. \cite{ParticleDataGroup:2024cfk} & Experiments & $<3.5$ MeV& $ < 3.8$ MeV\\
			Ref. \cite{Faessler:2007gv} & $DK$ & &$79.3\pm32.6$ KeV \\
			Ref. \cite{Yue:2025wcl} & $DK$ & &63.0$\sim$209 KeV\\
			Ref. \cite{Fu:2021wde} & $D^{(*)}K$ & $111 \pm15$ KeV& $132 \pm 7$ KeV\\
			Ref. \cite{Wu:2014era} & $c\bar s$ & $9.0 \pm 2.1$ KeV& $9.2\pm 2.3$ KeV\\
			Ref. \cite{Fajfer:2015zma} & $c\bar s$ & $2.7$$\sim$$3.4$ KeV& $2.4$$\sim$$4.7$ KeV\\
			Ref. \cite{Liu:2006jx} & $c\bar s$ & $1.86$$\sim$$4.42$ KeV& $3.68$$\sim$$8.71$ KeV\\
			Ref. \cite{Nielsen:2005zr} & tetraquark & & $6\pm2$ KeV\\
			Ref. \cite{Colangelo:2003vg} & $c\bar s$ & $7\pm 1$ KeV& $7\pm 1$ KeV\\
			Ref. \cite{Wei:2005ag} & $c\bar s$ & $35$$\sim$$51$ KeV& $34$$\sim$$44$ KeV\\
			Ref. \cite{Ishida:2003gu} & $c\bar s$  & $150\pm70$ KeV& $150\pm70$ KeV\\
			This work & $D^{(*)}K$ & ${24.8_{-5.4}^{+5.5}}$ KeV& ${10.3_{-2.9}^{+3.0}}$ KeV
		\end{tabular}\label{decay width}
	\end{ruledtabular}
\end{table}
In the present work, $\eta-{\pi^0}$ mechanism is considered to analyze the decay processes of $D_{s0}^{*}\to {D_s}\pi^0 $, ${D_{s1}}\to D_s^*\pi^0 $. It was discussed in Ref. \cite{Faessler:2007gv,Fu:2021wde,Yue:2025wcl} that the hadronic molecules can also decay through meson loops. It is one more isospin violating mechanism which can also lead to the same magnitude of decay width as $\eta-{\pi^0}$ mechanism. Thus, it will be an interesting work to study the decay processes with these two mechanism considered in the frame work of QCDSR. In addition, we predict two large decay widths of $T_{BK}\to BK$ and $T_{B^*K}\to B^*K$, which may be helpful for searching for these hadronic molecules in experiments.
		
\section{Conclusion}\label{sec5}

In this work, we perform the analysis of strong vertices of hadronic molecules $DK$, $D^{*}K$, $DK^{*}$, $BK$, $B^{*}K$, and $BK^{*}$ by using two methods within the framework of three-point QCDSR. According to comparison, the results obtained from these two methods are roughly consistent with each other. However, we sometime can not obtain stable sum rules for method II because the contributions of excited and continuum states of initial particle are neglected. These coupling constants are important input parameters in analyzing the decaying and producing processes of these charm/bottom strange states. Using the predicted coupling constants of method I, we also study the partial decay widths of $T_{D^{(*)}K}\to D_{s}^{(*)}\pi$ and $T_{B^{(*)}K}\to B^{(*)}K$.
According to comparison, we find that the experimental data of $D_{s0}^*(2317)$ and $D_{s1}(2460)$ can be well reproduced by our results. Thus, we temporarily support explaining these two states as $DK$, $D^{*}K$ hadronic molecules. Certainly, this conclusion needs to be confirmed by more theoretical analysis and experimental data. As for $BK$ and $B^{*}K$, we hope our calculating results can provide useful information for searching these hadronic states in the future in experiments.
\begin{center}
\begin{Large}
Acknowledgments
\end{Large}
\end{center}

This project is supported by National Natural Science Foundation, Grant Number 12575083 and Natural Science Foundation of HeBei Province, Grant Number A2018502124.		

\begin{widetext}
\appendix
\section{The convergence of OPE in the Borel platform}\label{appendix A}
\begin{figure*}[h!]
\centering
\includegraphics[width=18cm, trim=70 530 60 50, clip]{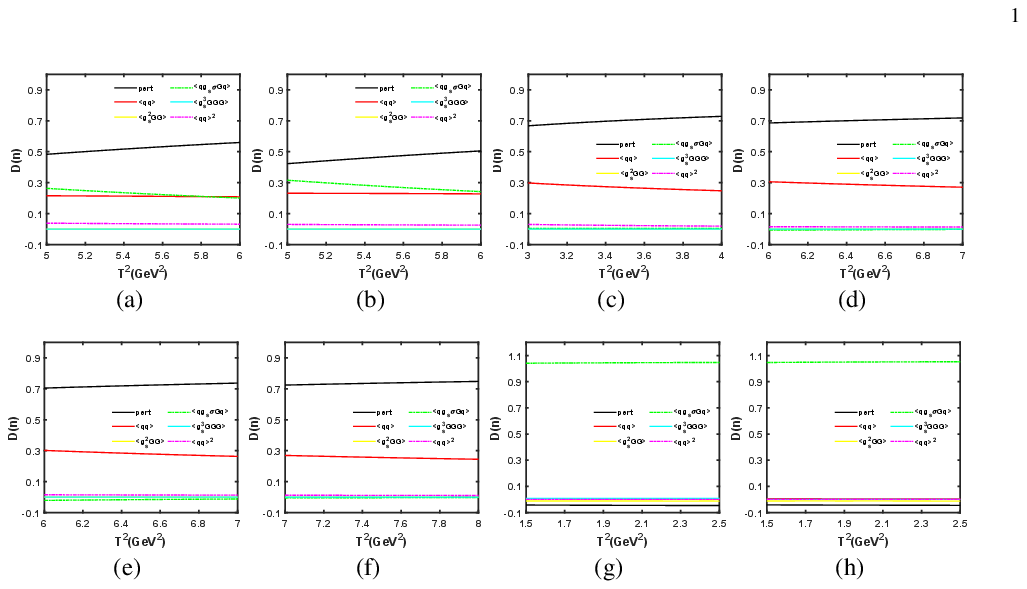}
\caption{The contribution of perturbative and different vacuum condensate terms vary with the Borel parameter.}\label{contribution of dimension}
\end{figure*}
\FloatBarrier
\end{widetext}

\end{document}